\documentclass[nofootinbib,superscriptaddress,longbibliography,prd,twocolumn,amsmath,amssymb,preprintnumbers]{revtex4-1}
\usepackage{bm}
\usepackage{graphicx}
\usepackage{hyperref}
\usepackage{color}
\preprint{LA-UR-20-30504}
\newcommand{\figscale}{0.6}

\newcommand{\gA}{g_\text{A}}
\newcommand{\cV}[1]{c_\text{V}^{#1}}
\newcommand{\cA}[1]{c_\text{A}^{#1}}
\newcommand{\thetaW}{\theta_\text{W}}

\newcommand{\pfrac}[2]{\left(\frac{#1}{#2}\right)}

\newcommand{\sfH}{\mathsf{H}}
\newcommand{\sfM}{\mathsf{M}}
\newcommand{\sfPi}{\mathsf{\Pi}}
\newcommand{\sfone}{\mathsf{1}}
\newcommand{\sfOmega}{\mathsf{\Omega}}
\newcommand{\sfLambda}{\mathsf{\Lambda}}
\newcommand{\sfV}{\mathsf{V}}
\newcommand{\sfC}{\mathsf{C}}

\newcommand{\vcp}{\protect{\vec{p}}}
\newcommand{\vcr}{\protect{\vec{r}}}
\newcommand{\htv}{\protect{\hat{v}}}

\newcommand{\bfP}{\mathbf{P}}
\newcommand{\bfB}{\mathbf{B}}
\newcommand{\bfe}{\mathbf{e}}
\newcommand{\bsigma}{\bm{\sigma}}

\newcommand{\Pa}{\bar{P}}
\newcommand{\bfPa}{\bar{\bfP}}
\newcommand{\Sa}{\bar{S}}

\newcommand{\rmi}{\mathrm{i}}
\newcommand{\rmd}{\mathrm{d}}

\newcommand{\gain}{\text{gain}}
\newcommand{\loss}{\text{loss}}
\newcommand{\Ris}{R_\text{IS}}

\newcommand{\nue}{\protect{\nu_e}}
\newcommand{\nut}{\protect{\nu_\tau}}
\newcommand{\anue}{\protect{\bar\nu_e}}
\newcommand{\anut}{\protect{\bar\nu_\tau}}

\newcommand{\thetav}{\theta_\text{v}}

\newcommand{\GF}{G_{\text{F}}}

\newcommand*{\UNM}{Department of Physics \& Astronomy, University of New Mexico, Albuquerque, New Mexico 87131, USA}

\newcommand*{\LANL}{Theoretical Division, Los Alamos National Laboratory, Los Alamos, New Mexico 87545, USA}

\begin{document}

\title{Fast flavor oscillations in dense neutrino media with collisions}

\author{Joshua D. Martin}
\email{jdmartin@lanl.gov}
\affiliation{{\UNM}}
\affiliation{{\LANL}}

\author{J. Carlson}
\affiliation{{\LANL}}

\author{Vincenzo Cirigliano}
\affiliation{{\LANL}}

\author{Huaiyu Duan}
\affiliation{{\UNM}}

\date{\today}

\begin{abstract}
We investigate the impact of the nonzero neutrino splitting and elastic neutrino-nucleon collisions on fast neutrino oscillations. Our calculations confirm that a small neutrino mass splitting and the neutrino mass hierarchy have very little effect on fast oscillation waves. We also demonstrate explicitly that fast oscillations remain largely unaffected for the time/distance scales that are much smaller than the neutrino mean free path but are damped on larger scales. This damping originates from both the direct modification of the dispersion relation of the oscillation waves in the neutrino medium and the flattening of the neutrino angular distributions over time. Our work suggests that fast neutrino oscillation waves produced near the neutrino sphere can propagate essentially unimpeded which may have ramifications in various aspects of the supernova physics.
\end{abstract}

\maketitle

\section{Introduction} \label{sec:intro}

Neutrino flavor oscillation is a quantum phenomenon caused by the misalignment of the weak-interaction states of the neutrinos in which they are produced and the eigenstates of their Hamiltonians (see, e.g., Ref.~\cite{Zyla:2020zbs} for a review). This phenomenon becomes very rich and interesting in, e.g., a core-collapse supernova or a neutron star merger, where a large portion of a dense neutrino medium can experience flavor oscillations collectively because of the coupling among the neutrinos themselves (see, e.g., Ref.~\cite{Duan:2010bg} for a review). Because of its nonlinear nature, collective flavor oscillations in dense neutrino media pose a great challenge to both numerical and analytic approaches and have yet to be fully understood. Nevertheless, important progress has been made in recent years which has shed light on this intriguing phenomenon.

Early studies of collective neutrino oscillations have employed stringent symmetric conditions such as the homogeneity and isotropy in the early universe (see, e.g., Refs.~\cite{Kostelecky:1993yt,Abazajian:2002qx}), and the time invariance, spherical symmetry, and axial symmetry in supernovae (see, e.g., Refs.~\cite{Pastor:2002we,Duan:2006jv}). However, these symmetries are likely to be broken spontaneously by neutrino oscillations because of the flavor instabilities in the neutrino media even if these symmetries are present initially \cite{Raffelt:2013rqa,Duan:2013kba,Duan:2014gfa,Abbar:2015fwa,Dasgupta:2015iia} (see Ref.~\cite{Chakraborty:2016yeg} for a review). The breaking of these symmetries have been confirmed by numerical calculations with simplified models \cite{Mirizzi:2013wda,Mirizzi:2015hwa,Mirizzi:2015fva,Martin:2019kgi,Martin:2019dof}.

Collective oscillations occur on time or distance scales of $(\omega \mu)^{-1/2}$ when neutrinos of different species have the same angular distribution \cite{Kostelecky:1994dt}, where $\omega$ and $\mu$ are the vacuum oscillation frequency of the neutrino and the strength of the neutrino-neutrino coupling, respectively. Much attention has been paid recently to the fast oscillations that occur on the scales of $\mu^{-1}$ when the neutrino species have different angular distributions (see, e.g., Refs.~\cite{Sawyer:2015dsa,Chakraborty:2016lct,Dasgupta:2016dbv}; see also Ref.~\cite{Tamborra:2020cul} for a review). These fast oscillations can occur in the region where neutrinos decouple from the matter and, therefore, may have a great impact on the supernova physics (see, e.g., Refs.~\cite{Abbar:2019zoq,Glas:2019ijo}).

Since its introduction by Ref.~\cite{Banerjee:2011fj}, the stability analysis of the linearized equations of motion has been used extensively to predict the prospect of flavor oscillations in static or homogeneous neutrino gases. For dynamic inhomogeneous oscillations, one can study the analytical structures of the dispersion relations of the neutrino media which distinguish various kinds flavor instabilities \cite{Izaguirre:2016gsx,Airen:2018nvp,Yi:2019hrp,Capozzi:2019lso}. The validity of this approach has been verified in numerical simulations in both the linear and nonlinear regimes \cite{Dasgupta:2018ulw,Martin:2019gxb}.

In an earlier study we have demonstrated that fast oscillation waves can spontaneously appear in collisionless dense neutrino media under suitable conditions and redistribute the electron lepton number (ELN) as the flavor waves propagate in space \cite{Martin:2019gxb}. (The calculations in Ref.~\cite{Bhattacharyya:2020dhu} obtained flavor-depolarized steady states instead of wave-like solutions. However, it appears that the simulation tools used in Ref.~\cite{Bhattacharyya:2020dhu} have difficulty in maintaining the causality over a long time scale.) It has been shown that neutrino collisions, although rare outside the neutrino sphere of a supernova, can influence the results of collective oscillations \cite{Cherry:2012zw,Cirigliano:2018rst}. Inside the supernova core, collisions may even trigger fast oscillations \cite{Capozzi:2018clo,Shalgar:2020wcx} which in turn will affect the supernova dynamics, nucleosynthesis, and neutrino signals.

In this work we investigate the impacts of neutrino collisions on the formation and propagation of the fast oscillation waves. As the first step, we include only the neutrino-nucleon scattering as in Ref.~\cite{Cirigliano:2017hmk}. We will also investigate the validity of the zero mass-splitting approximation for fast oscillations which has been widely adopted in the literature but was questioned in Ref.~\cite{Shalgar:2020xns}. The rest of the paper is organized as follows. After establishing the mathematical formalism and the physical model (Sec.~\ref{sec:formalism}), we demonstrate the impacts of the collisions on fast neutrino oscillations both analytically (Sec.~\ref{sec:analytics}) and numerically (Sec.~\ref{sec:numerics}). We conclude by summarizing our main results and discussing their implications (Sec.~\ref{sec:discussion}).

\section{Formalism and Model} \label{sec:formalism}

\subsection{General description}

We use the flavor density matrix $\rho_\vcp(t,\vcr)$ to describe the flavor content of the momentum mode $\vcp$ of a neutrino medium at the spacetime point $(t,\vcr)$, where its diagonal elements are the occupation numbers in the corresponding weak-interaction states, and the off-diagonal elements are the coherences between these states \cite{Sigl:1992fn}. In this work we employ the two-flavor mixing between $\nue$ and $\nut$, where $\nut$ represents a linear combination of the physical $\mu$ and $\tau$ flavor neutrinos. (See, however, Ref.~\cite{Capozzi:2020kge} for a possible limitation of the two-flavor mixing approximation.) The flavor density matrix obeys the following quantum kinetic equation in the mean field limit
\cite{Sigl:1992fn, Vlasenko:2013fja}
\begin{align}
  (\partial_t + \htv\cdot\vec\nabla) \rho_\vcp = -\rmi [\sfH_\vcp, \rho_\vcp]  + \sfC_\vcp,
  \label{eq:QKE}
\end{align} 
where the Hamiltonian $\sfH_\vcp(t, \vcr)$ dictates the coherent flavor evolution of the neutrino medium, and $\sfC_\vcp(t, \vcr)$ determines the incoherent evolution due to the collisions, absorptions and emissions of the neutrinos. Here we have adopted the natural units with $\hbar=c=1$, and we have assumed that the neutrinos are ultra-relativistic so that they have velocities $\htv=\vcp/|\vcp|$. We have ignored the gravitational effect and the possible oscillations between the neutrino and the antineutrino or the sterile neutrino. The flavor density matrix $\bar\rho_\vcp(t, \vcr)$ for the antineutrino is defined in a similar way and obeys a similar equation of motion.

The Hamiltonian in Eq.~\eqref{eq:QKE} has three components:
\begin{align}
  \sfH_\vcp = \sfOmega_\varepsilon + \sfLambda + \sfV_\htv .
  \label{eq:H}
\end{align}
The first component of the Hamiltonian, $\sfOmega_\varepsilon = \sfM^2/2\varepsilon$, describes the vacuum oscillation of the neutrino,  where $\sfM^2$ and $\varepsilon=|\vcp|$ are the mass-squared matrix and the energy of the neutrino, respectively. The second component of the Hamiltonian is the matter potential $\sfLambda = \sqrt2\GF \text{diag}[n_e, 0]$, where $\GF$ is the Fermi constant, and $n_e$ the net number density of the electron \cite{Wolfenstein:1977ue}. Here we have assumed that the matter does not have a (significant) overall motion and the number densities of the heavy leptons are negligible. The last component of the Hamiltonian is due to the neutrino-neutrino coupling \cite{Fuller:1987aa,Notzold:1987ik,Pantaleone:1992xh} and takes the form of 
\begin{align}
  \sfV_\htv = \sqrt2\GF
  \int(1-\htv\cdot\htv')(\rho_\vcp - \bar\rho_\vcp)
  \,\frac{\rmd^3 p'}{(2\pi)^3}.
\end{align}
The Hamiltonian for the antineutrino is similar except with $\sfOmega_\varepsilon$ replaced by $-\sfOmega_\varepsilon$.

For the incoherent evolution we consider only the elastic collisions of the neutrinos and antineutrinos off non-relativistic nucleons in this work: 
\[\nu + N \rightarrow \nu + N,\] 
where $\nu$ represents a neutrino or antineutrino of any flavor, and $N$ can be either a neutron ($n$) or proton ($p$). This collision effect is given by \cite{Blaschke:2016xxt}
\begin{align}
  \sfC_\vcp = 
  \frac{1}{2}\{ \sfPi^\gain_\vcp, \sfone - \rho_\vcp\} 
  -\frac{1}{2}\{ \sfPi^\loss_\vcp, \rho_\vcp\},  
  \label{eq:coll0}
\end{align}
where $\{\cdot,\cdot\}$ denotes the anti-commutator. In the limit that the neutrino energy $\varepsilon$ is much less than the nucleon mass $m_N$, the energies of the neutrino before and after the collision are approximately equal. In this isoenergetic limit, the gain and loss potentials become
\begin{subequations}
  \begin{align}
    \sfPi^\gain_\vcp &= \int\!\Ris(\htv, \htv')\rho_{\vcp'} 
    \,\frac{\rmd^3 p'}{(2\pi)^3} 
    \nonumber\\ 
    & = \frac{\varepsilon^2}{(2\pi)^3} 
    \int\!\Ris^0(\htv, \htv')\rho_{\vcp'}\, \rmd\Omega_{\htv'}, \\
    \sfPi^\loss_\vcp &= \int\!\Ris(\htv, \htv') (\sfone-\rho_{\vcp'})
    \,\frac{\rmd^3 p'}{(2\pi)^3}
    \nonumber\\ 
    & = \frac{\varepsilon^2}{(2\pi)^3} 
    \int\!\Ris^0(\htv, \htv')(\sfone -\rho_{\vcp'})
    \, \rmd\Omega_{\htv'},
  \end{align}    
\end{subequations}
where $\Ris(\htv, \htv') = \Ris^0(\htv, \htv')\delta(\varepsilon' - \varepsilon)$ is the isoenergetic scattering kernel, and $\rmd\Omega_{\htv'}$ is the differential solid angle pointing in the direction of $\htv'$. For non-relativistic, non-degenerate nucleons, one has  \cite{Bruenn:1985en}
\begin{align}
  \Ris^0(\htv, \htv') = 2\pi \GF^2 \sum_N 
  &n_N
  \{[(\cV{N})^2 +3  (\cA{N})^2] 
  \nonumber\\
  & + 
  [(\cV{N})^2 - (\cA{N})^2](\htv\cdot\htv')\},
\end{align} 
where $n_N$ ($N=n,p$) are the number densities of the nucleons,
\begin{subequations}
  \begin{align}
    \cV{p} &= \frac{1}{2} - 2\sin^2\thetaW, &
    \cA{p} &= \frac{\gA}{2}, \\
    \cV{n} &=  -\frac{1}{2}, &
    \cA{n} &= -\frac{\gA}{2} 
  \end{align}
\end{subequations}
are the weak coupling constants with $\gA\approx 1.27$, and $\sin^2\thetaW\approx 0.23$. For isoenergetic scatterings, Eq.~\eqref{eq:coll0} is reduced to
\begin{align}
    \sfC_\vcp = \frac{\varepsilon^2}{(2\pi)^3} 
    \int\!\Ris^0(\htv, \htv')(\rho_{\vcp'} -\rho_\vcp)
    \, \rmd\Omega_{\htv'}.
\end{align} 

\subsection{Homogeneous axial neutrino gas}

We investigate the fast neutrino oscillations on very short time and distance scales over which the matter distribution can be considered as both constant and uniform. As in Ref.~\cite{Martin:2019gxb}, we simplify the problem by imposing the translation symmetries along both the $x$ and $y$ directions and the axial symmetry about the $z$ axis. We further assume that the neutrino field is nearly uniform along $z$ at $t=0$ but can change later on. Because the isoenergetic collisions do not change the energies of the neutrinos, we adopt the single-energy approximation by assuming that the evolution of the neutrino gas is represented by monochromatic neutrinos and antineutrinos of a characteristic energy $\varepsilon$. This single-energy approximation can be removed in future works to study the impact of inelastic collisions.

For this simplified model, it is convenient to express the flavor density matrices in terms of the polarization vectors,
\begin{subequations}
  \label{eq:P}
  \begin{align}
    \rho_\vcp(t,\vcr) & \propto \left(\frac{F_+}{2}\right) P_{u,0}(t,z) + \left(\frac{F_-}{2}\right) \bsigma\cdot\bfP_u(t,z), \\
    \bar\rho_\vcp(t,\vcr) &\propto \left(\frac{\bar{F}_+}{2}\right) \Pa_{u,0}(t,z) +  \left(\frac{\bar{F}_-}{2}\right)  \bsigma\cdot\bfPa_u(t,z),
  \end{align}    
\end{subequations}
where $u$ is the velocity component of the neutrino along the $z$ axis, and
$\sigma_i$ ($i=1,2,3$) are the Pauli matrices. The trace term $P_{u,0}$ ($\Pa_{u,0}$) describes the total spatial and angular distribution of the neutrino (antineutrino) of both flavors, while the polarization vector $\bfP_u$ ($\bfPa_u$) describes the coherent flavor distribution of the neutrino (antineutrino). We normalize both the trace terms and polarization vectors at $t=0$ and $z=0$ so that
\begin{align}
    &\quad\int_{-1}^1 P_{u,0}(0, 0)\,\rmd u =
    \int_{-1}^1 \Pa_{u,0}(0, 0)\,\rmd u \nonumber\\
    &= \int_{-1}^1 P_{u,3}(0, 0)\,\rmd u =
    \int_{-1}^1 \Pa_{u,3}(0, 0)\,\rmd u = 1,
    \label{eq:P-norm}
\end{align}
and
\begin{subequations}
  \begin{align}
    F_\pm &= n_\nue(0,0) \pm n_\nut(0,0), \\
    \bar{F}_\pm &= n_\anue(0,0) \pm n_\anut(0,0),
  \end{align}    
\end{subequations}
where $n_\nu(t,z)$ is the number density of the neutrino species $\nu$.

In the representation of the polarization vectors, one has
\begin{align}
  -\rmi[\sfH_\vcp, \rho_\vcp]
  &\longrightarrow 
  \Big[\mu\int_{-1}^1(1-u u')(\bfP_{u'} - \alpha \bfPa_{u'})\rmd u'
  \nonumber\\
  &\qquad + (\bfB + \lambda \bfe_3)\Big]
  \times\bfP_{u},
\end{align}
where 
\begin{align}
  \alpha= \frac{F_-}{\bar{F}_-},
\end{align}
and
\begin{align}
  \bfB &=\pfrac{\delta m^2}{2\varepsilon}[\sin(2\thetav)\bfe_1 - \cos(2\thetav)\bfe_3],  \\
  \lambda &=\sqrt2\GF n_e, \\
  \intertext{and} 
  \mu &=\sqrt2\GF F_-
\end{align}
measure the strengths of the vacuum, matter, and neutrino potentials, respectively. Here $\delta m^2$ and $\thetav$ are the mass-squared difference and the vacuum mixing angle of the neutrino, respectively, and $\bfe_i$ ($i=1,2,3$) are the unit basis vectors in the flavor space. We will work in the reference frame that rotates about $\bfe_3$ in flavor space in which
\begin{align}
  \bfB + \lambda\bfe_3\longrightarrow -\omega\bfe_3,
\end{align}
where the effective oscillation frequency $\omega=(\delta m^2/2\varepsilon)\cos(2\thetav)$ is positive for the normal neutrino mass hierarchy (NH) and negative for the inverted hierarchy (IH). The values of $P_{u,3}$ and $\Pa_{u,3}$, which determine the flavor transformation of the neutrino and antineutrino, are unaffected by the rotating frame transformation. In this rotating frame, Eq.~\eqref{eq:QKE} becomes
\begin{widetext}
  \begin{subequations}
    \label{eq:eom}
    \begin{align}
      (\partial_t + u \partial_z) P_{u,0} 
      &= -\kappa_0 P_{u,0} + 
      \frac{1}{2}\int_{-1}^1\left(\kappa_0 - \frac{\kappa_1}{3} u u'\right)P_{u',0}\, \rmd u',
      \label{eq:P0-eom}\\
      (\partial_t + u \partial_z) \bfP_{u}
      &= \left[-\omega\bfe_3 + \mu\int_{-1}^1(1-u u')(\bfP_{u'} - \alpha\bfPa_{u'})\, \rmd u' \right] \times\bfP_{u}  
      -\kappa_0 \bfP_{u}
      + \frac{1}{2}\int_{-1}^1\left(\kappa_0 - \frac{\kappa_1}{3} u u'\right)\bfP_{u'}\, \rmd u',
      \label{eq:P-eom}\\
      (\partial_t + u \partial_z) \Pa_{u,0} 
      &= -\kappa_0 \Pa_{u,0} + 
      \frac{1}{2}\int_{-1}^1\left(\kappa_0 - \frac{\kappa_1}{3} u u'\right)
      \Pa_{u',0}\,\rmd u',\\
      (\partial_t + u \partial_z) \bfPa_{u}
      &= \left[+\omega\bfe_3 + \mu\int_{-1}^1(1-u u')(\bfP_{u'} -\alpha \bfPa_{u'})\, \rmd u' \right] \times\bfPa_{u}  
      -\kappa_0 \bfPa_{u}
      + \frac{1}{2}\int_{-1}^1\left(\kappa_0 - \frac{\kappa_1}{3} u u'\right)\bfPa_{u'}\,\rmd u',
      \label{eq:Pa-eom}
    \end{align}
  \end{subequations}
  \end{widetext}
where 
\begin{subequations}
  \begin{align}
    \kappa_0 &= \frac{3}{\pi}\GF^2\varepsilon^2\sum_N n_N \left[(\cA{N})^2 + \frac{(\cV{N})^2}{3} \right] ,\\
    \kappa_1 &= \frac{3}{\pi}\GF^2\varepsilon^2\sum_N n_N [(\cA{N})^2 - (\cV{N})^2  ] 
  \end{align}
\end{subequations}
are the positive constants that measure the strength of the isoenergetic scattering of the neutrinos by the nucleons. For a charge neutral matter consisting of free nucleons and electrons, one obtains 
\begin{align}
  \frac{\kappa_1}{\kappa_0} \approx 0.5
\end{align}  
for the electron fraction $Y_e=0.3$. We will use this nominal ratio in the rest of the paper. 

\section{Collisional damping}\label{sec:analytics}
The scattering of the neutrinos off a homogeneous thermalized matter tend to make the neutrino gas more homogeneous and isotropic, and thus damps the neutrino oscillations. This effect can be understood analytically in our model.

\subsection{Total neutrino angular distributions}
\begin{figure}[htb]
  \begin{center}
    $\begin{array}{c}
      \includegraphics*[scale=\figscale]{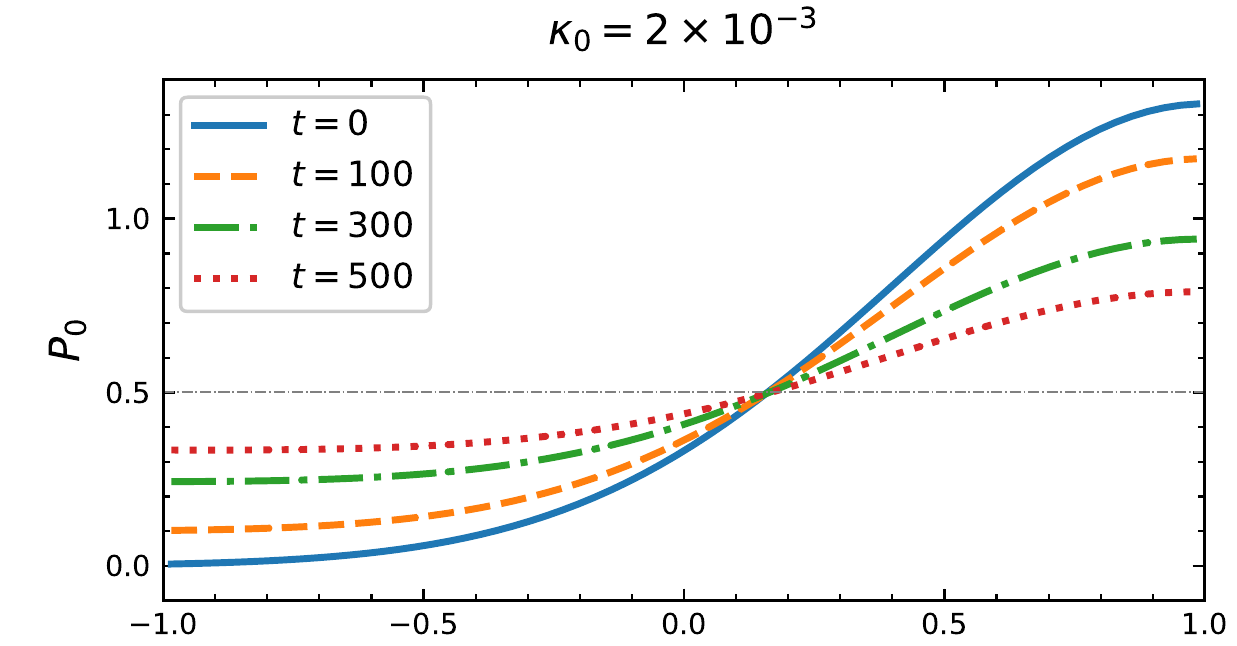} \\
      \includegraphics*[scale=\figscale]{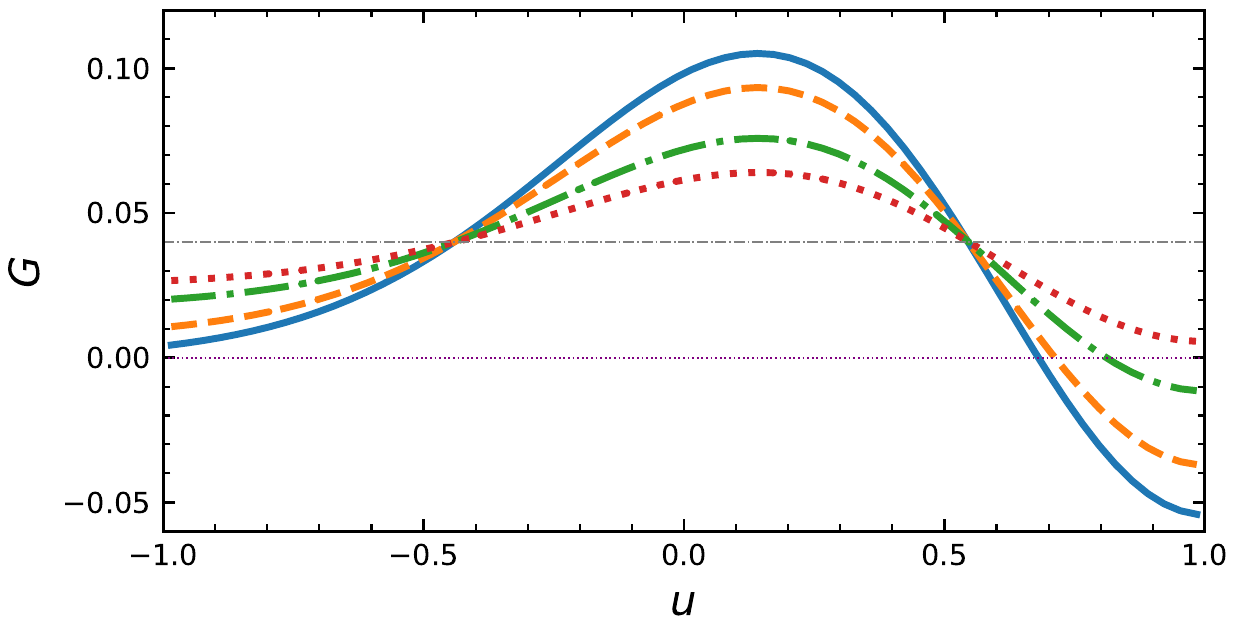} 
      \end{array}$
  \end{center}
  \caption{The overall neutrino distribution $P_{u,0}$ (top panel) and the ELN distribution $G_u$ (bottom panel) of a homogeneous gas at a few times (as indicated in the legend) in the strong collision scenario with $\kappa_0=2\times10^{-3}$. The thin horizontal dot-dashed lines in both panels are the asymptotic limits when the distributions are fully isotropic. A necessary condition for fast flavor instabilities to exist is that the ELN distribution in the bottom panel crosses the dotted line along $G_u=0$.}
  \label{fig:damping}
\end{figure}

We first consider the effect of the collisions on the evolution of the total density distributions $P_{u,0}$ and $\Pa_{u,0}$ of a homogeneous neutrino gas. It is obvious from Eq.\eqref{eq:eom} that these total distributions are decoupled from the evolution of the flavor polarization vectors. We expand the total neutrino distribution in terms of the Legendre polynomials,
\begin{align}
  P_{u,0}(t) = \sum_{\ell=0}^\infty f_\ell(t) L_\ell(u),
\end{align}
where $L_\ell(u)$ is the Legendre polynomial of degree $\ell$ with $L_0(u) = 1$ and $L_1(u)=u$. Using Eq.~\eqref{eq:P0-eom} and the orthogonal relation
\begin{align}
  \int_{-1}^1 L_\ell(u) L_{\ell'}(u) \,\rmd u = \frac{2}{2\ell+1}\,\delta_{\ell'\ell},
\end{align}
we obtain
\begin{subequations}
  \begin{align}
    \dot{f}_0 &= 0,\\
    \dot{f}_1 &= -(\kappa_0 + \kappa_1) f_1, \\
    \dot{f}_\ell &= -\kappa_0 f_\ell \quad (\ell\geq 2),
  \end{align}
\end{subequations}
where we have dropped the spatial derivative for the homogeneous gas. The above equations have the simple solutions
\begin{subequations}
  \begin{align}
    f_0(t) &= \text{const.},  \\
    f_1(t) &\propto e^{-(\kappa_0+\kappa_1)t}, \\
    f_\ell(t) &\propto e^{-\kappa_0 t} \quad(\ell\geq 2).
  \end{align}
\end{subequations}
Because both $\kappa_0$ and $\kappa_1$ are positive, all the multipoles decay away on the time scale of $\kappa_0^{-1}$ except for the monopole ($\ell=0$) which remains constant. The overall distribution of the antineutrino behaves in a similar way.

As a concrete example, we consider the following homogeneous distribution at $t=0$:
\begin{align}
  P_{u,0}(0) = g(u, 0.6)
\end{align}
where
\begin{align}
  g(u, w)\propto e^{-(u-1)^2/2w^2}
\end{align}
with the normalization condition $\int_{-1}^1 g(u,w)\,\rmd u=1$. In the upper panel of Fig.~\ref{fig:damping} we show a few snapshots of $P_{u,0}(t)$ for the scenario with $\kappa_0/\mu=2\times10^{-3}$. Following many other works on fast oscillations, we measure all energies in terms of $\mu$ and distance/time in $\mu^{-1}$ by setting 
\begin{align}
  \mu = 1.
\end{align}
Fig.~\ref{fig:damping} clearly shows that the collisions make the neutrino gas more isotropic over the timescale of $\kappa_0^{-1}$.

\subsection{Flavor instabilities}\label{sec:instabilities}

\begin{figure*}[htb!]
  \begin{center}
    $\begin{array}{@{}l@{\hspace{0.1in}}l@{}}
      \includegraphics*[scale=\figscale]{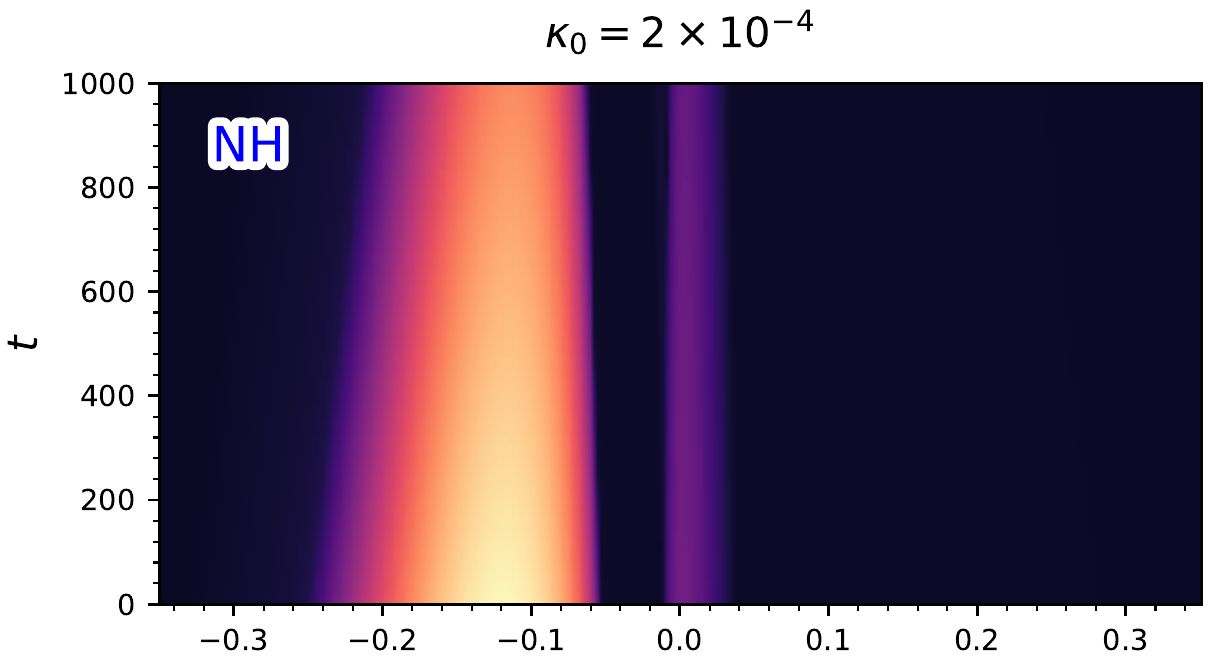} &
      \includegraphics*[scale=\figscale]{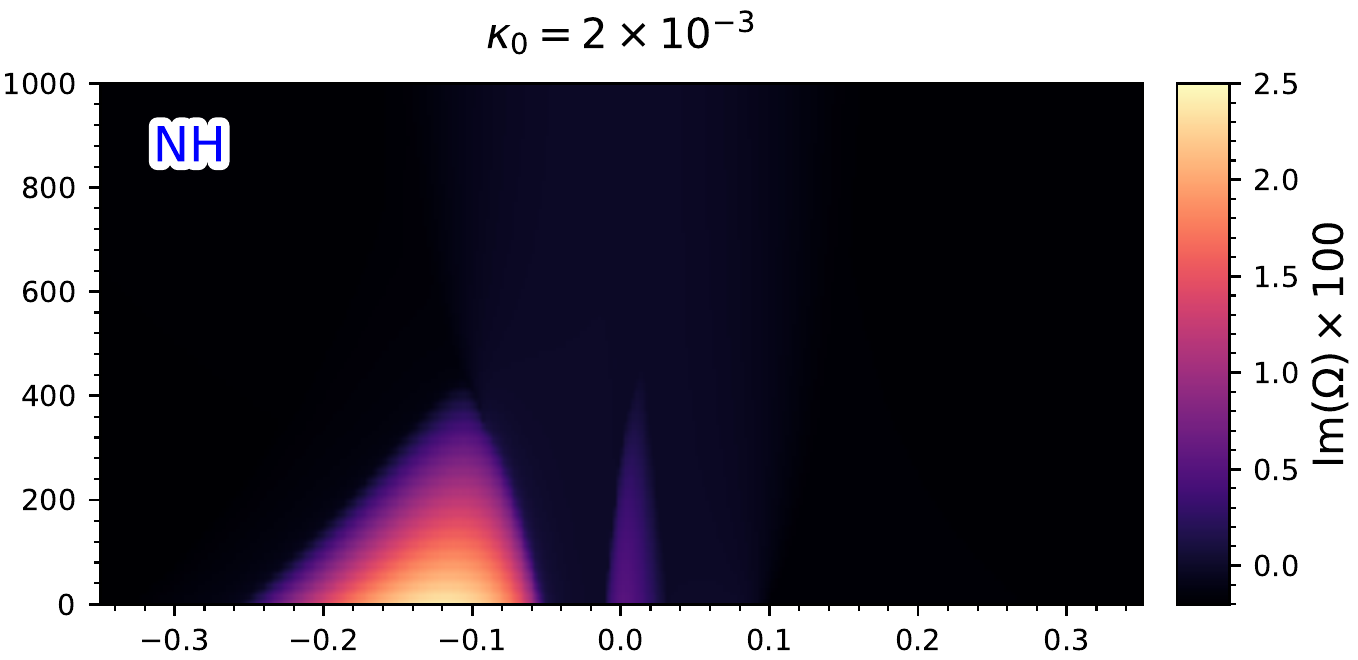} \\
      \includegraphics*[scale=\figscale]{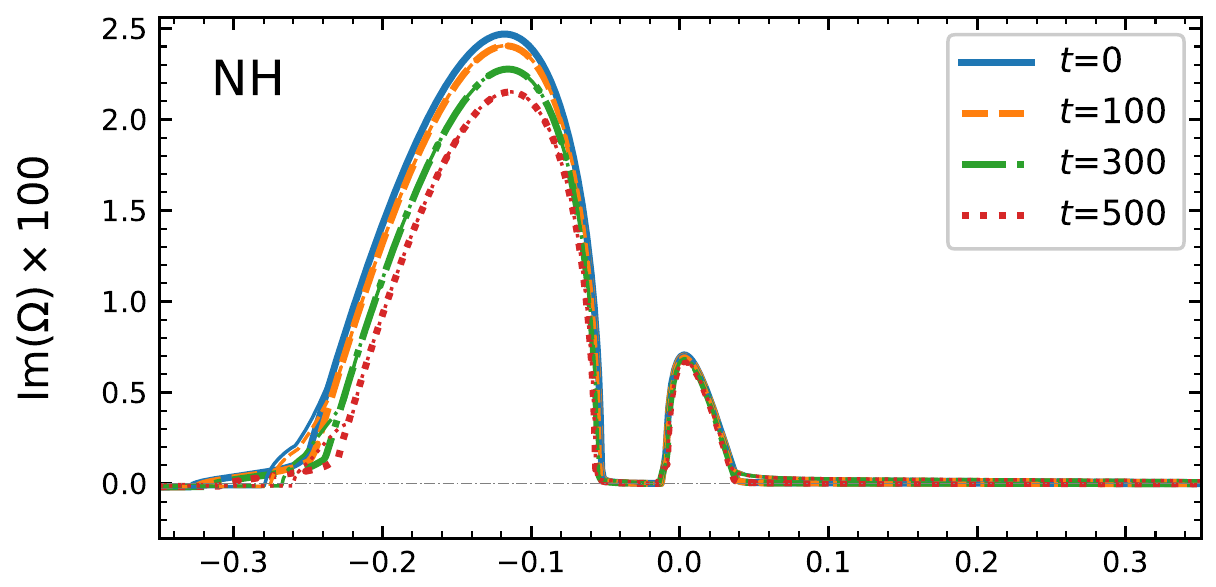} &
      \includegraphics*[scale=\figscale]{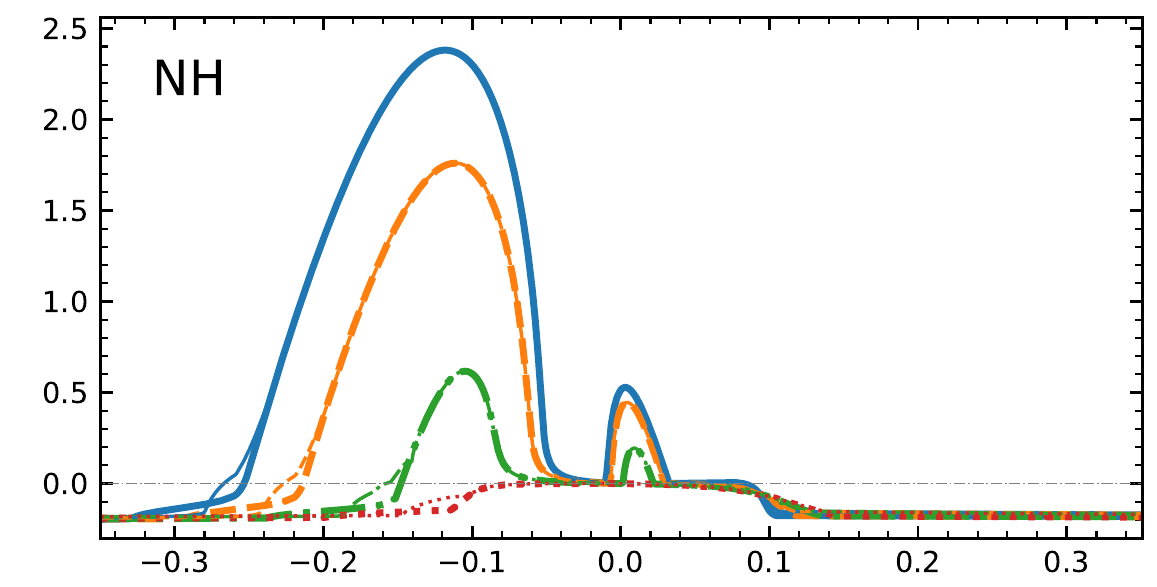} \\
      \includegraphics*[scale=\figscale]{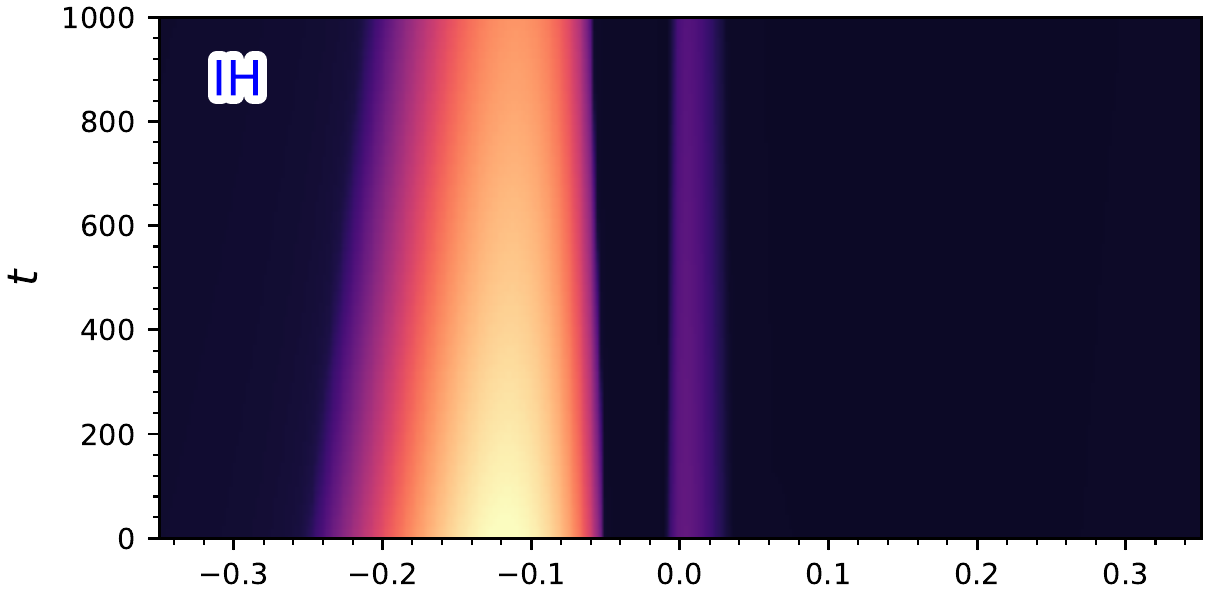} &
      \includegraphics*[scale=\figscale]{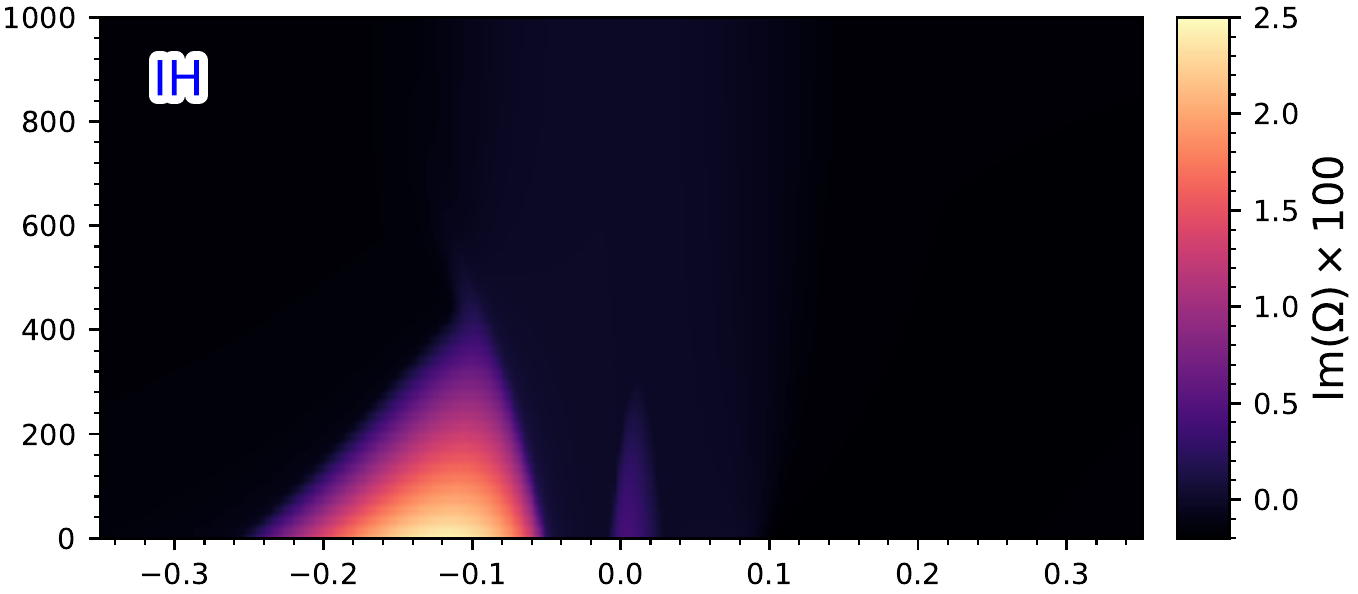} \\
      \includegraphics*[scale=\figscale]{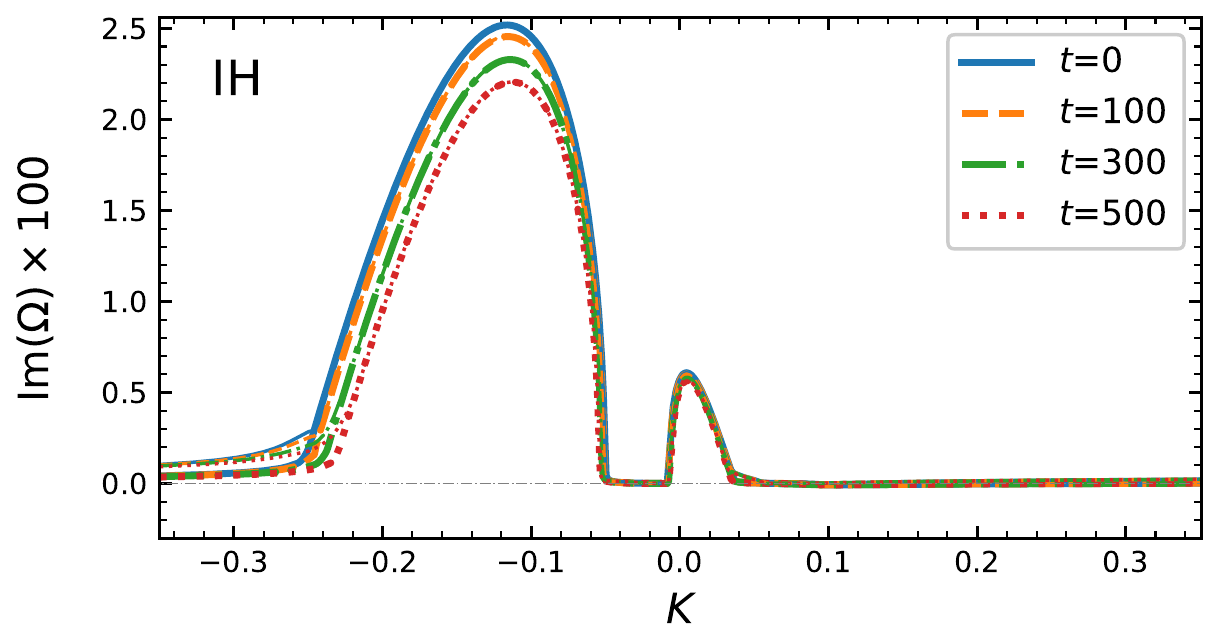} &
      \includegraphics*[scale=\figscale]{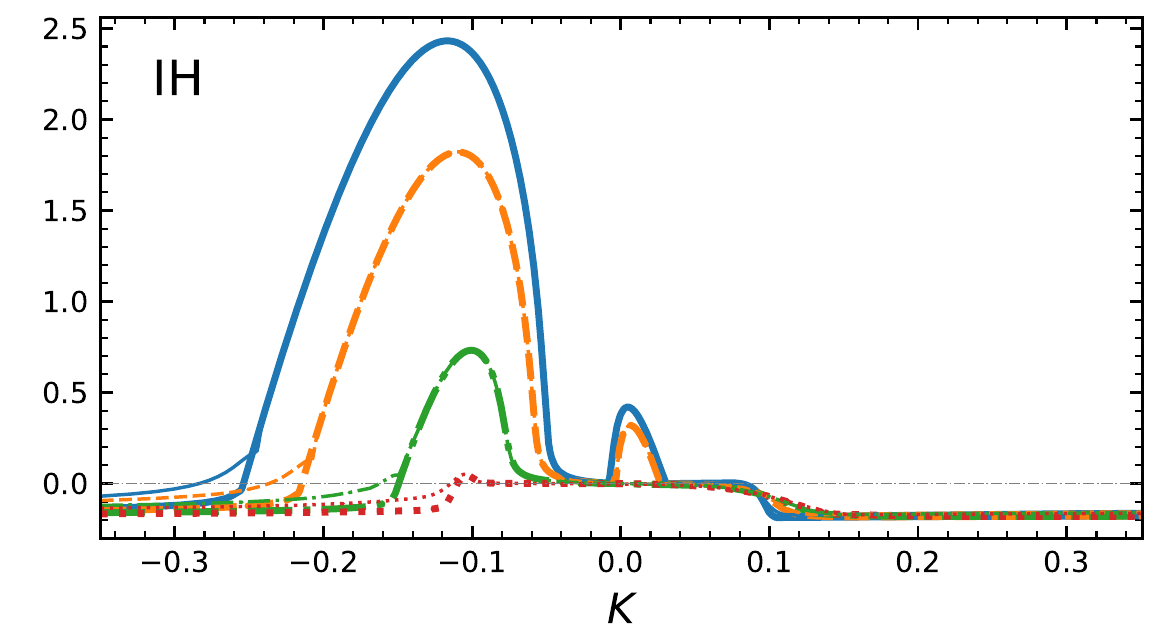} 
    \end{array}$
  \end{center}
  \caption{The time evolution of the instabilities in a weak collision scenario ($\kappa_0=2\times10^{-4}$, left panels) and a strong collision scenario ($\kappa_0=2\times10^{-3}$, right panels) with the normal neutrino mass hierarchy (NH, upper two rows) and the inverted hierarchy (IH, lower two rows), respectively. The color of each pixel in the figures of the first and third rows represents the maximum exponential growth rate $\text{Im}(\Omega)$ (in units of $\mu/100$) of the normal mode with wave number $K$ (horizontal axis) at time $t$ (vertical axis). The same growth rates as functions of $K$ at a few times (as indicated in the legend) are plotted in the the second and fourth rows. A value below the thin dot-dashed line along $\text{Im}(\Omega)=0$ in these panels represents a exponential decay instead of growth of the wave amplitude. The thin curves in the second and fourth rows are computed with 64 angle bins, and the rest are computed with 256 angle bins.
  }\label{fig:DR}
\end{figure*}

A dense, collisionless neutrino gas can support collective oscillation modes or normal modes with $S_u, \Sa_u \propto e^{\rmi(K z -\Omega t)}$, where
\begin{align}
S_u = P_{u,1} - \rmi P_{u,2}
\quad\text{and}\quad
\Sa_u = \Pa_{u,1} - \rmi \Pa_{u,2},  
\end{align}  
are the amplitudes of the waves, and $\Omega$ and $K$ are the frequency and wave number of the normal modes, respectively. Using Eq.~\eqref{eq:eom} we obtain
\begin{widetext}
  \begin{subequations}
    \label{eq:S-t}
    \begin{align}
      \rmi(\partial_t + u\partial_z)S_u 
        &= 
        (-\omega - \rmi\kappa_0) S_u
        +\mu \int_{-1}^1(1-u u') [G_{u'} S_u - P_{u,3}(S_{u'}-\alpha \Sa_{u'})]\,\rmd u'
         + \frac{\rmi}{2}\int_{-1}^1 
         \left(\kappa_0 - \frac{\kappa_1}{3} u u'\right) S_{u'}\,\rmd u',\\
      \rmi(\partial_t + u\partial_z)\Sa_u 
        &= 
        (+\omega - \rmi\kappa_0) \Sa_u
        +\mu \int_{-1}^1(1-u u') [G_{u'} \Sa_u - \Pa_{u,3} (S_{u'}-\alpha\Sa_{u'})]\,\rmd u'
         + \frac{\rmi}{2}\int_{-1}^1 
         \left(\kappa_0 - \frac{\kappa_1}{3} u u'\right) \Sa_{u'}\,\rmd u',
    \end{align}  
  \end{subequations}
\end{widetext}
where
\begin{align}
  G_u(t, z) &= P_{u,3}(t, z) - \alpha \Pa_{u,3}(t, z)
  \label{eq:G}
\end{align}
is the instantaneous ELN distribution. From Eq.~\eqref{eq:S-t} one can derive the dispersion relation $\Omega(K)$ of the neutrino oscillation wave in the linear regime when $S_u$ and $\Sa_u$ are small \cite{Izaguirre:2016gsx}. As in Ref.~\cite{Martin:2019gxb}, we will consider only the branches of the dispersion relation that have real wave numbers. For these branches, a positive (negative) value of the imaginary component of the collective oscillation frequency $\text{Im}(\Omega)$ gives the exponential growth (decay) rate of the corresponding normal mode.

The impact of neutrino collisions are manifested in Eq.~\eqref{eq:S-t} both directly and indirectly. The direct impact is the presence of $\kappa_0$ and $\kappa_1$ in this equation which tends to decrease the overall value of $\text{Im}(\Omega)$. The indirect impact is on the ELN distribution $G_u(t)$ which changes over time because of the collisions. 

When $S_u$ and $\Sa_u$ are small, $G_u(t,z)=G_u(t)$ remains homogeneous, and its time evolution is governed by the following equation which is also derived from Eq.~\eqref{eq:eom}:
\begin{align}
  \dot{G}_u 
      \approx -\kappa_0 G_u + 
      \frac{1}{2}\int_{-1}^1 \left(\kappa_0 - \frac{\kappa_1}{3} u u'\right)
       G_{u'}\,\rmd u'.
    \label{eq:G-t}
\end{align}
Because this equation is similar to Eq.~\eqref{eq:P0-eom} without the spatial derivative, we expect the ELN distribution to flatten out on the timescale of $\kappa_0^{-1}$ as $P_{u,0}$ does. As an example, we consider a homogeneous neutrino gas with 
\begin{subequations}
  \begin{align}
    \bfP_u(t=0,z) &= g(u, 0.6)\, \bfe_3, \\
    \bfPa_u(t=0,z) &= g(u, 0.53)\, \bfe_3,
  \end{align}    
\end{subequations}
and $\alpha=0.92$. We chose this initial condition because it produces the $G_\text{4b}$ spectrum in Ref.~\cite{Martin:2019gxb} against the calculations of which we will make comparisons. We computed the evolution of $G_u(t)$ with $\kappa_0=2\times10^{-3}$, and the results are shown in the lower panel of Fig.~\ref{fig:damping}. As expected, the ELN distribution becomes isotropic over time.

To further investigate the impact of neutrino collisions on the flavor instabilities of the neutrino gas, we computed $P_u(t)$ and $\Pa_u(t)$ in both a strong collision scenario with $\kappa_0=2\times10^{-3}$ and a weak collision scenario with $\kappa_0=2\times10^{-4}$. Both $P_u(t)$ and $\Pa_u(t)$ obey the same equation of motion as $G_u(t)$ in Eq.~\eqref{eq:G-t}. We then computed the dispersion relation $\Omega(K)$ from Eq.~\eqref{eq:S-t} as a function of time in both collision scenarios. In these calculations, we assume an effective vacuum oscillation frequency $\omega=\pm10^{-5}$, where the plus and minus signs are for the NH and IH, respectively. The maximum exponential growth rates $\text{Im}(\Omega)$ of the normal modes for all scenarios are shown in Fig.~\ref{fig:DR}. We have employed both 256 and 64 angle bins in these calculations. These numbers of angle bins are much smaller than those needed in the static models because the spurious instabilities are negligible in dynamic fast oscillations \cite{Martin:2019gxb}.

Fig.~\ref{fig:DR} shows that a nonzero mass splitting and the neutrino mass hierarchy have a very small effect on the dispersion relation when $|\omega|/\mu\ll 1$. It also shows that the presence of the collisions does reduce the instabilities. In fact, a wide range of the normal modes which are stable in the absence of collisions now have a decay rate of order $\kappa_0^{-1}$ at $t=0$. As time progresses, the maximum growth rates of the unstable modes also decrease as the ELN distribution flattens. In the strong collision scenario, the flavor instabilities virtually disappear at $t\gtrsim \kappa_0^{-1}$ when the ELN crossing vanishes (see the lower panel Fig.~\ref{fig:damping}).

\section{Numerical results} \label{sec:numerics}

To verify the damping effects of the collisions that are presented in the previous section, we carried out a suite of calculations using the numerical schemes similar to Ref.~\cite{Martin:2019gxb} but with collisions. We assume that the neutrino gas is confined within a periodic box of length $L$. We employ the same parameters as in Sec.~\ref{sec:instabilities} except for small initial perturbations to initial polarization vectors: 
\begin{subequations}
  \begin{align}
    \bfP_u(0, z) &= g(u, 0.6) \,[\epsilon(z)\,\bfe_1 +  \sqrt{1-\epsilon^2(z)} \,\bfe_3], \\
    \bfPa_u(0, z) &= g(u, 0.53) \, [\epsilon(z)\,\bfe_1 +  \sqrt{1-\epsilon^2(z)} \,\bfe_3],
  \end{align}
\end{subequations}
where
\begin{align}
  \epsilon(z) = \epsilon_0 e^{-(z-z_0)^2/50}
\end{align}
with $z_0=L/2$ and $\epsilon_0\ll1$ being a small positive constant. We employed 64 angle bins in all the calculations which reproduce the instabilities quite well (see Fig.~\ref{fig:DR}). Because the results with the NH and IH are very similar, the calculations presented in this section assume the NH, i.e.\ $\omega=10^{-5}$, unless otherwise noted. 

\subsection{Linear regime}
\begin{figure*}[htb!]
  \begin{center}
    $\begin{array}{@{}l@{\hspace{0.01in}}l@{\hspace{0.01in}}l@{}}
      \includegraphics*[scale=\figscale]{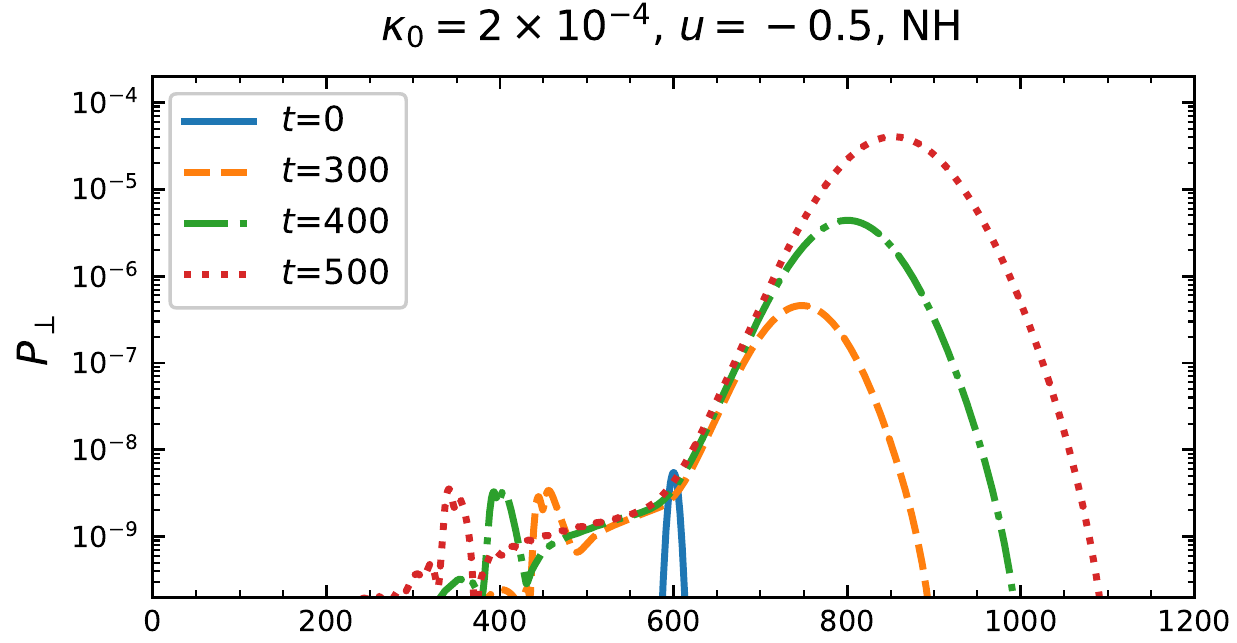} &
      \includegraphics*[scale=\figscale]{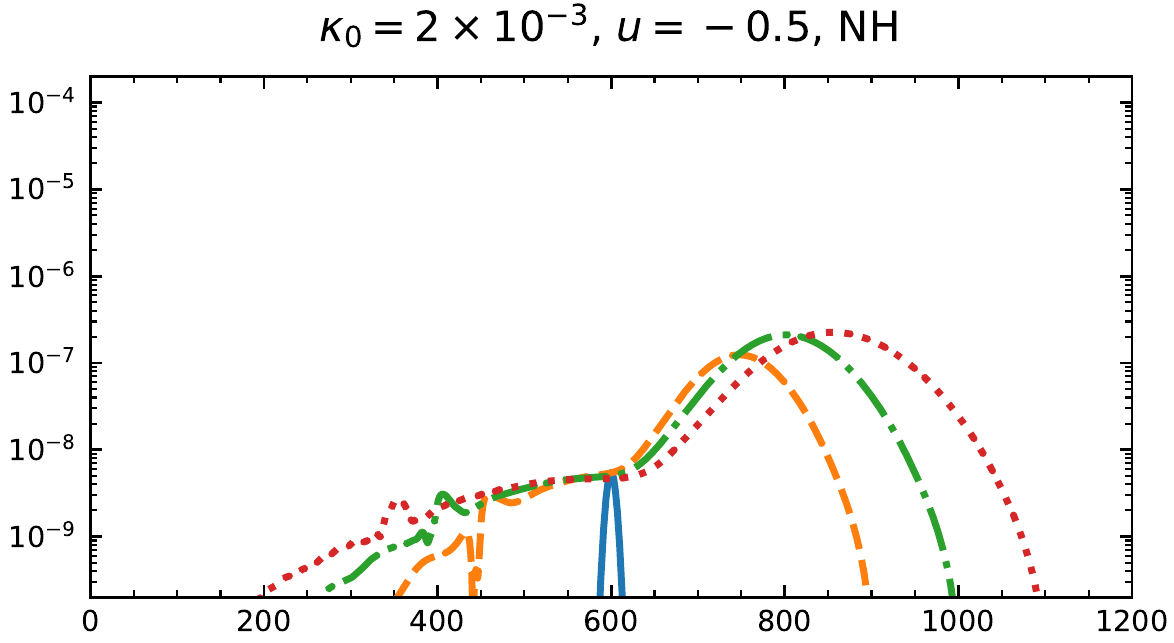} \\
      \includegraphics*[scale=\figscale]{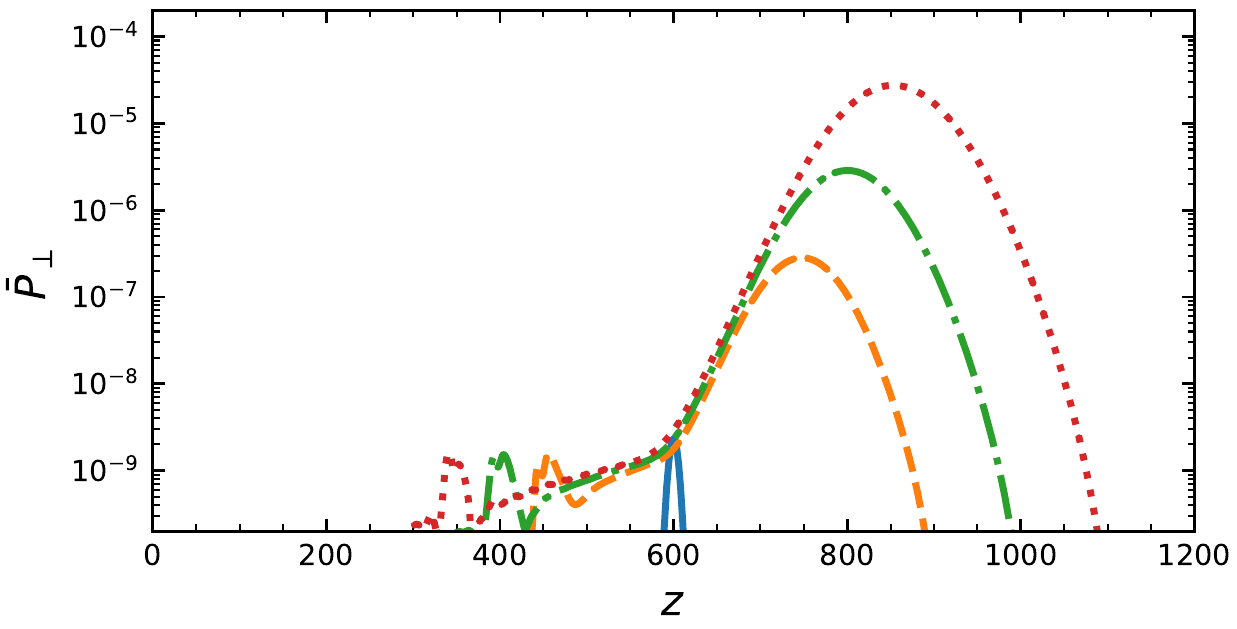} &
      \includegraphics*[scale=\figscale]{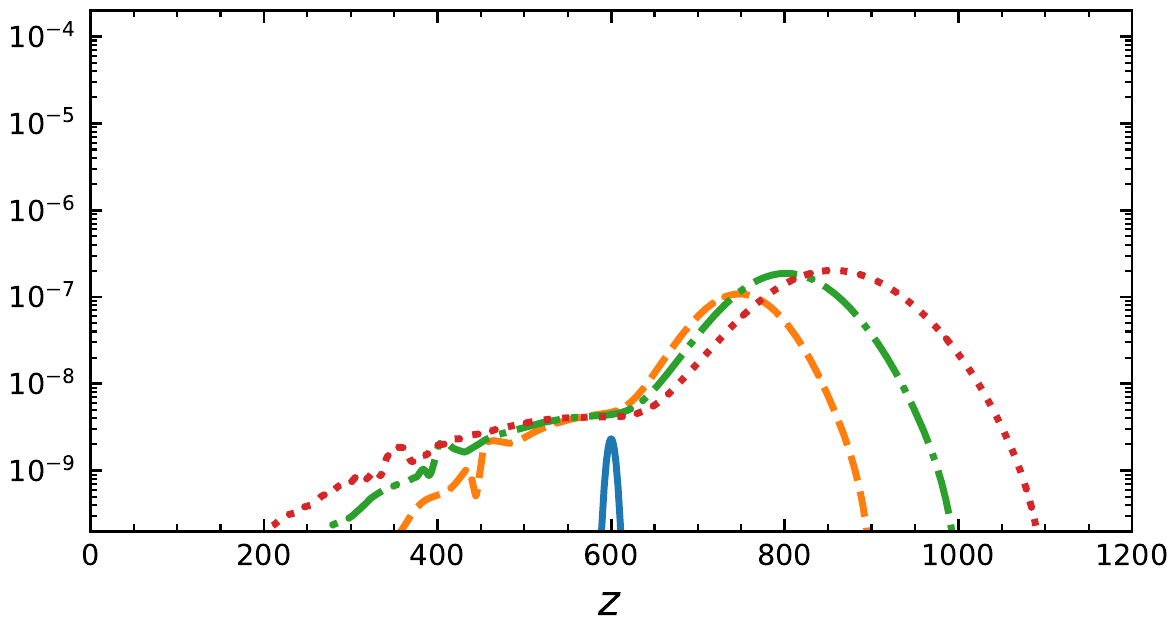} 
    \end{array}$
  \end{center}
  \caption{The time evolution of the transverse components of the polarization vectors of the neutrino and antineutrino, $P_\perp$ (top panels) and $\Pa_\perp$ (bottom panels), of the velocity mode $u\approx -0.5$ in a weak collision scenario (left panels) and a strong collision scenario (right panels), respectively. The normal neutrino mass hierarchy (NH) is employed in both calculations.
  }\label{fig:linear-regime}
\end{figure*}

We first consider the impact of the neutrino collisions in the linear regime where the transverse components of the polarization vectors,
\begin{align}
  P_{\perp,u} = |S_u|
  \qquad\text{and}\qquad
  \Pa_{\perp,u} = |\Sa_u|,
\end{align}
are always small. For this purpose we took $\epsilon_0 = 10^{-7}$ and  performed the calculations on $48000$ discrete spatial bins  equally spaced within a periodic box of size $L=1200$ which is about $1$ m for a typical value of $\mu\sim10^5\text{ km}^{-1}$ on the surface of the proto-neutron star. In Fig.~\ref{fig:linear-regime} we show the development of both $P_\perp$ and $\Pa_\perp$ of the velocity mode $u\approx -0.5$ in the weak and strong collision scenarios, respectively. The neutrinos and antineutrinos behave similarly in both scenarios, although there are small quantitative differences between $P_\perp$ and $\Pa_\perp$.

The results in the weak collision scenario is similar to the earlier result without the neutrino collisions. (See the right panel of Fig.~2 in Ref.~\cite{Martin:2019gxb}.) In this scenario, the initial localized perturbation splits into two peaks. One peak is transported to the left because of a (largely) real branch of the dispersion relation. The other peak grows exponentially while moving to the right because of a temporal instability. This instability is absolute because the regions of the exponential amplitude growth at later times always enclose the earlier regions \cite{Sturrock:1958zz}. This can be seen, for example, by noting that the right peaks of the dashed curves in the left panels of Fig.~\ref{fig:linear-regime} are fully enclosed by the dot-dashed curves which in turn are enclosed by the dotted curves. 

In the strong collision scenario, however, the left moving peaks are buried in extended envelopes which are caused by the scattering from other velocity modes. The right moving peaks grow fast initially because of the absolute instability, but the growth slows down as time progresses. We note that, unlike the weak collision scenario, the dot-dashed curves (at $t=400$) in the right panels of Fig.~\ref{fig:linear-regime} do not fully enclose the dashed curves ($t=300$). In other words, the wave amplitude decreases in the previously perturbed region after the instability has swept through it. This is a well-known feature of the convective instability \cite{Sturrock:1958zz}. It has been shown in Ref.~\cite{Yi:2019hrp} that an absolute instability can be weakened into a convective instability when the ELN crossing becomes shallower. As the ELN distribution is further flattened, the crossing eventually disappears and the instability vanishes. As a result, the exponential growth of the wave amplitude stops, which is indicated by the similar heights of the right peaks of the dot-dashed and dotted curves ($t=500$).

\subsection{Nonlinear regime}
\begin{figure*}[htb!]
  \begin{center}
    $\begin{array}{c c}
      \includegraphics*[scale=\figscale]{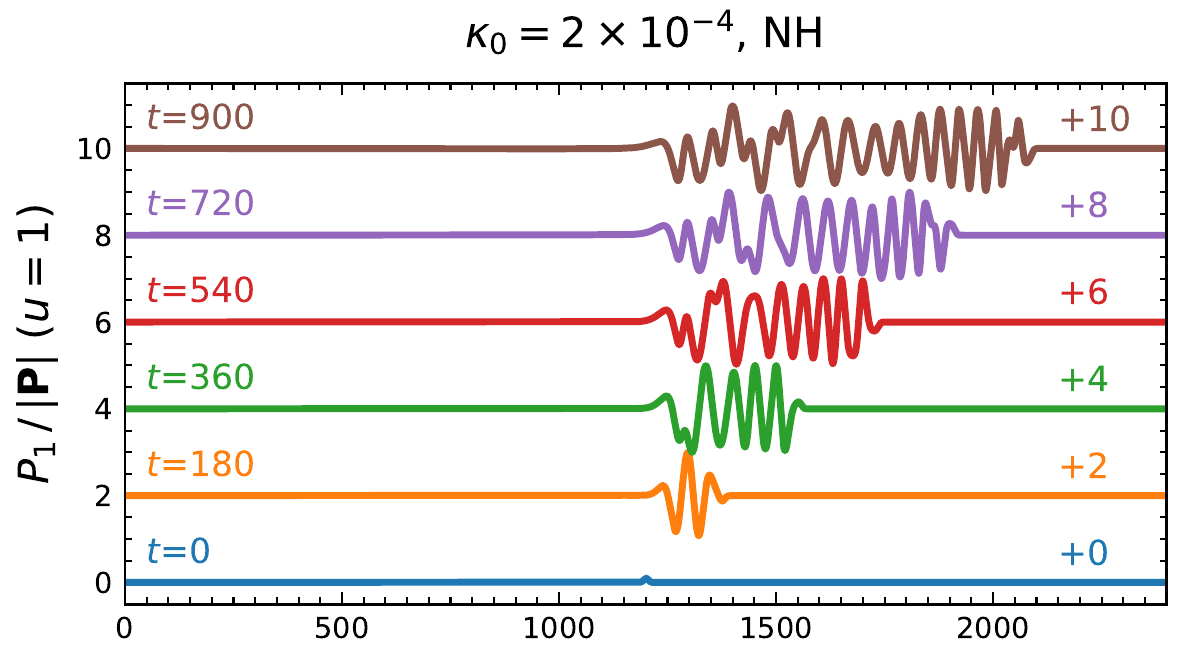} &
      \includegraphics*[scale=\figscale]{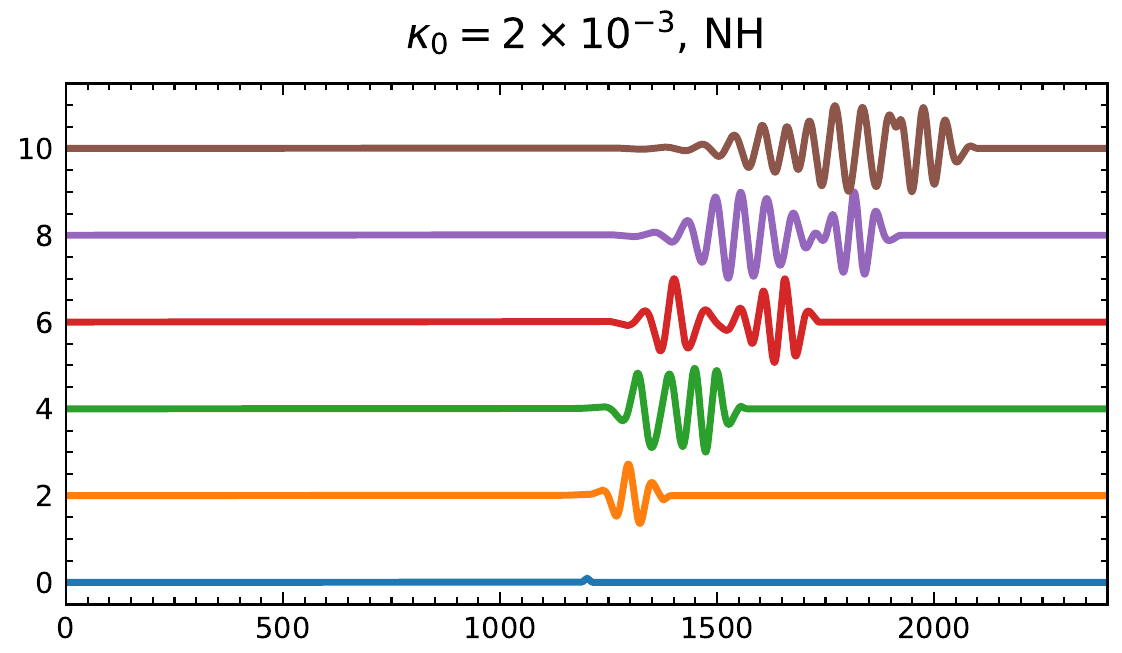} \\
      \includegraphics*[scale=\figscale]{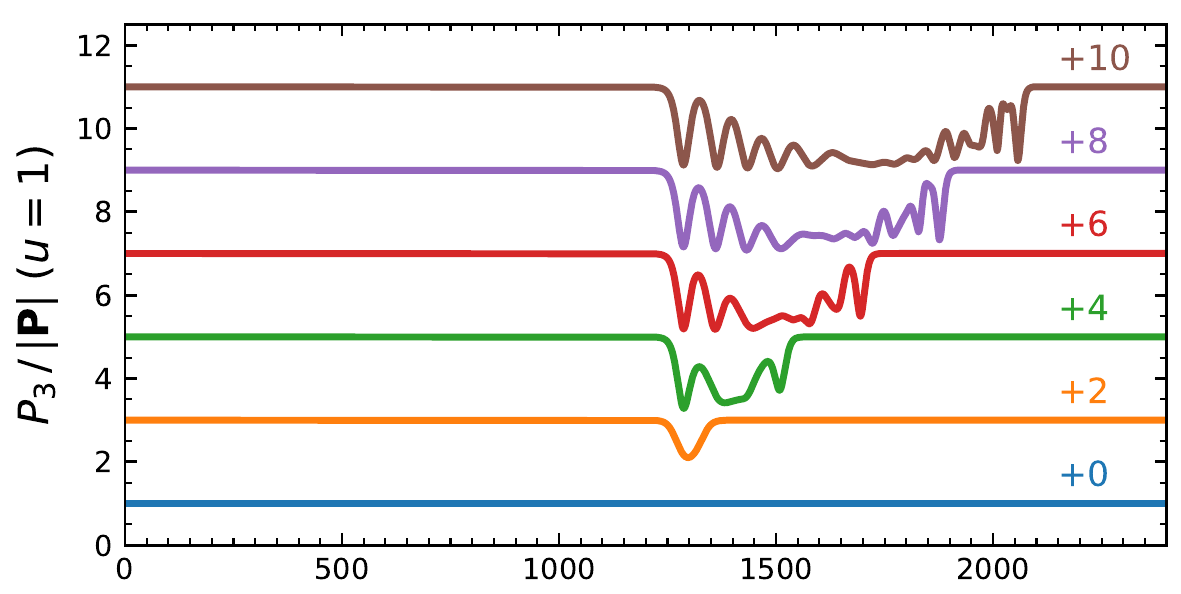} &
      \includegraphics*[scale=\figscale]{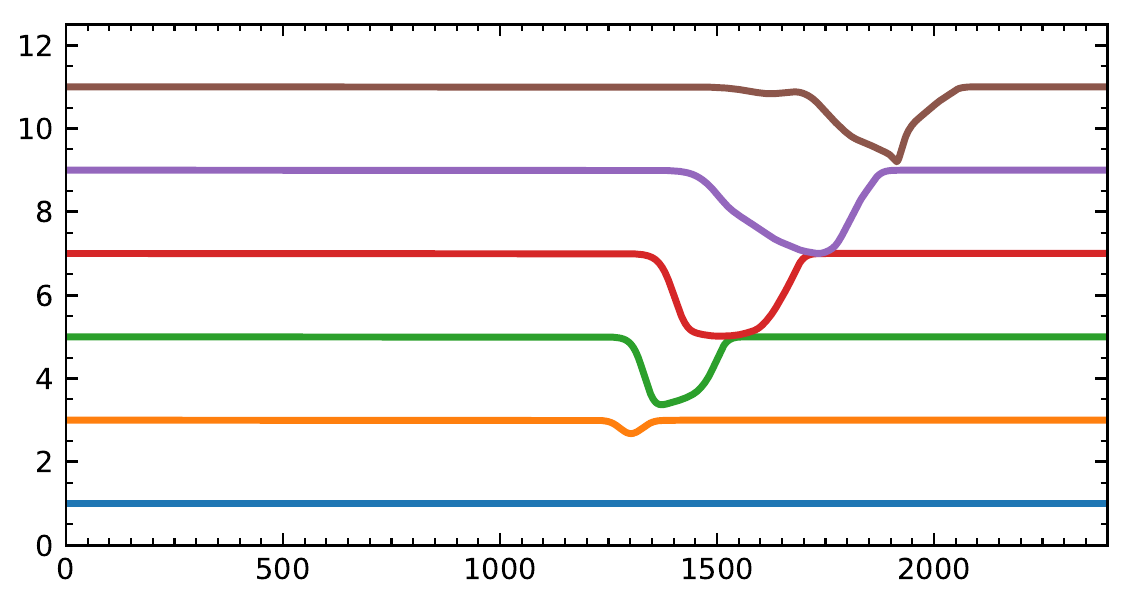} \\
      \includegraphics*[scale=\figscale]{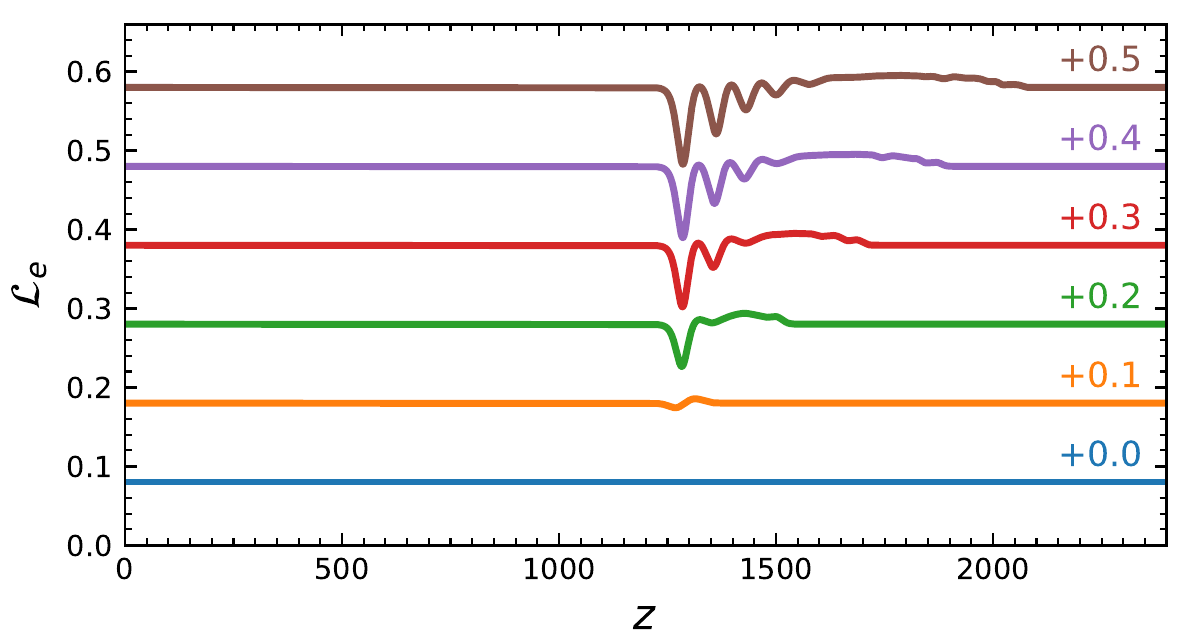} &
      \includegraphics*[scale=\figscale]{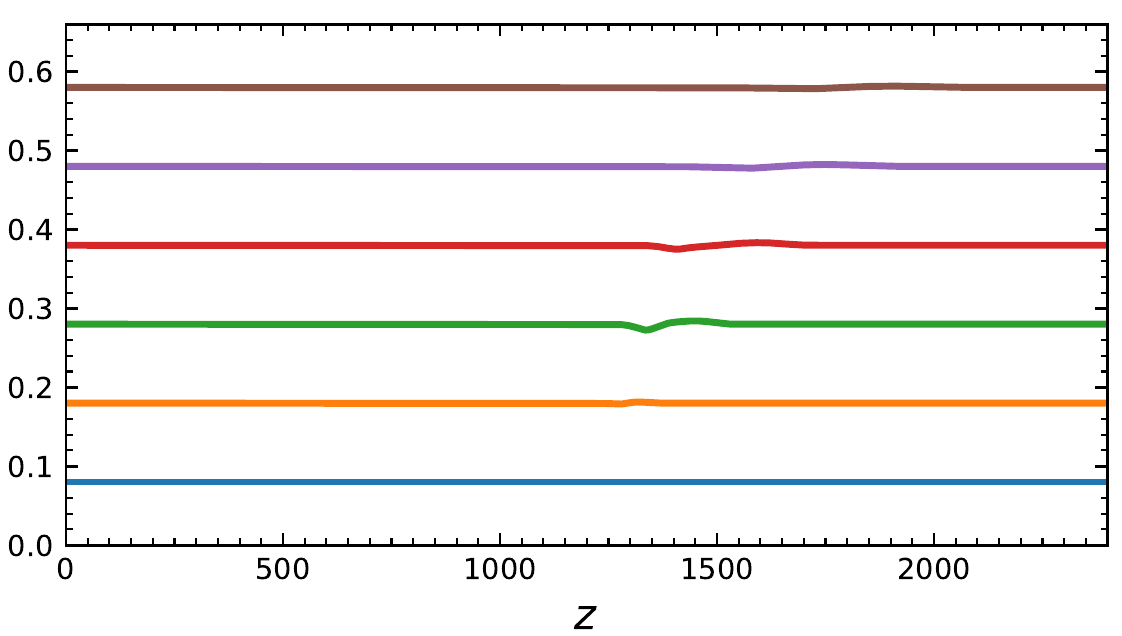} 
    \end{array}$
  \end{center}
  \caption{The polarization vector components $P_1$ (upper panels) and $P_3$ (middle panels), both normalized by the magnitude of the polarization vector $|\bfP|$ and for the velocity mode $u\approx 1$, and the electron lepton number (ELN) density $\mathcal{L}_e$ (lower panels) of a neutrino gas at various times $t$ (as labeled) as functions of the spatial coordinate $z$. The left and right panels are for the weak and strong collision scenarios discussed in the text. The normal neutrino mass hierarchy (NH) is employed in both calculations. The curves are offset from each other by 2 units in the upper two rows and by $0.1$ in the bottom row for clarity.
  }\label{fig:P-u1}
\end{figure*}

To investigate the impact of the collisions on fast oscillations in the nonlinear regime, we took $\epsilon_0=0.1$ and performed the calculations again on $48000$ equal spaced lattice points but in a larger box of size $L=2400$. 

In Fig.~\ref{fig:P-u1} we show several snapshots of the polarization vectors of the velocity mode $u\approx 1$ in the whole box in both the weak and strong collision scenarios, respectively. We also show the ELN densities
\begin{align}
  \mathcal{L}_e = \int_{-1}^1 G_u(t, z)\,\rmd u
\end{align}
in the same figure. The results in the weak collision scenario are again very similar to their counterparts in the calculation without collisions (see the right panels of Fig.~3 in Ref.~\cite{Martin:2019gxb}). The initial perturbation quickly grows into the nonlinear regime and spawns a flavor oscillation wave the range of which expands over time. These results also demonstrate the features of an absolute instability because the flavor conversion regions at earlier times are fully enclosed by those at later times. The results for the strong collision scenario are initially like those of a convective instability because the flavor conversion regions of later times are larger than but do not fully enclose those at earlier times. Because the ELN crossing has vanished at $t\gtrsim500$ in this scenario (see the lower panel of Fig.~\ref{fig:damping}), the flavor oscillation wave is dispersed according to the dispersion relation while propagating.

As in the neutrino gas without collisions, the ELN in the weak collision scenario is redistributed through the propagation of the neutrino oscillation wave. In contrast, the ELN density remains mostly homogeneous in the strong collision scenario for the reason to be explained below.

\begin{figure*}[htb!]
  \begin{center}
    $\begin{array}{@{}l@{\hspace{0.01in}}l@{\hspace{0.01in}}l@{}}
      \includegraphics*[scale=\figscale]{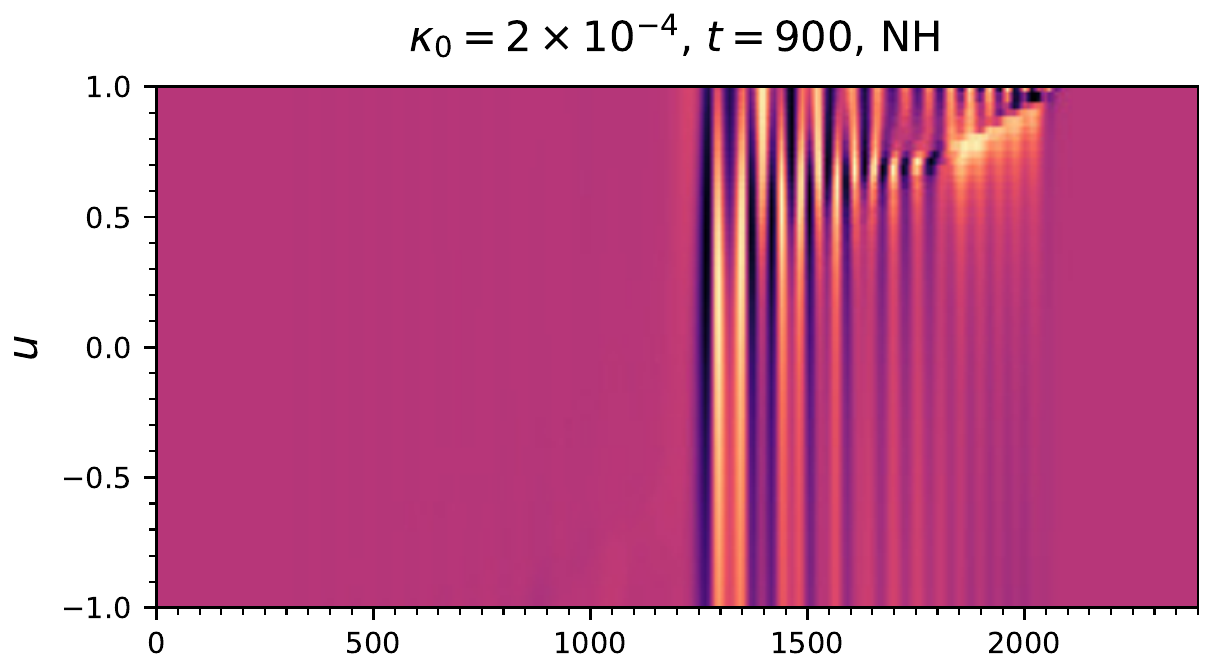} &
      \includegraphics*[scale=\figscale]{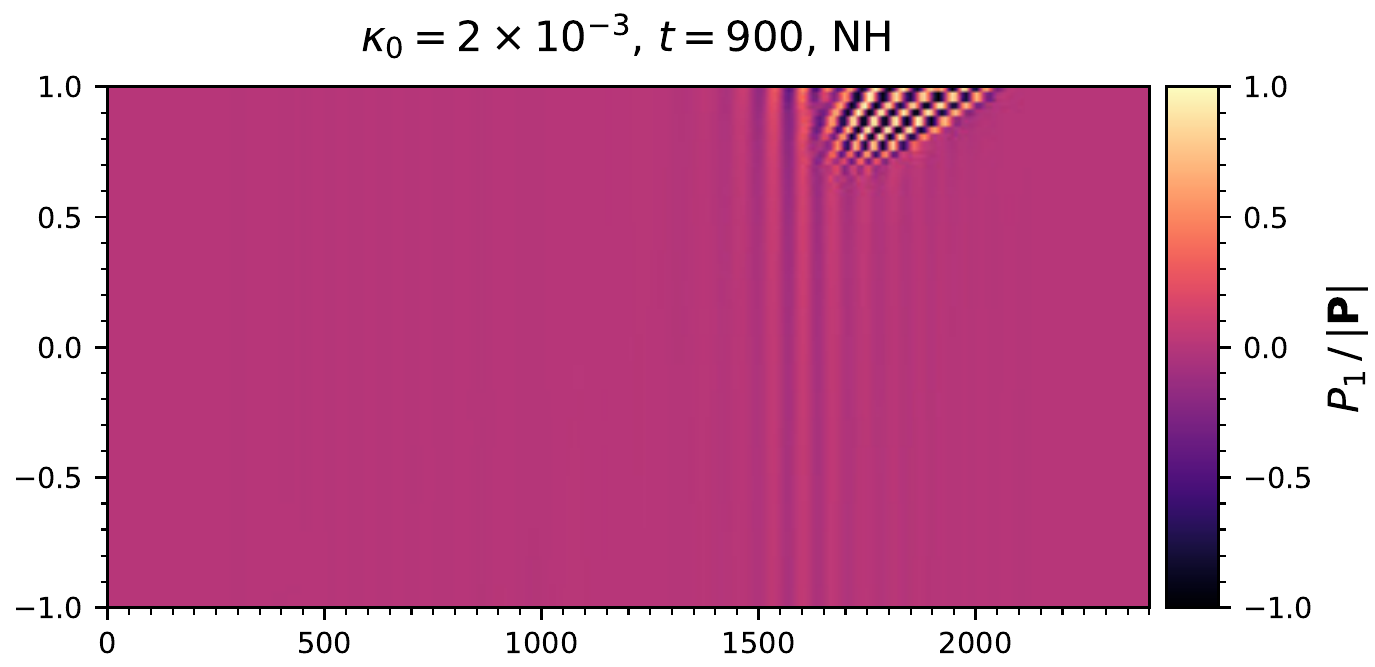} \\
      \includegraphics*[scale=\figscale]{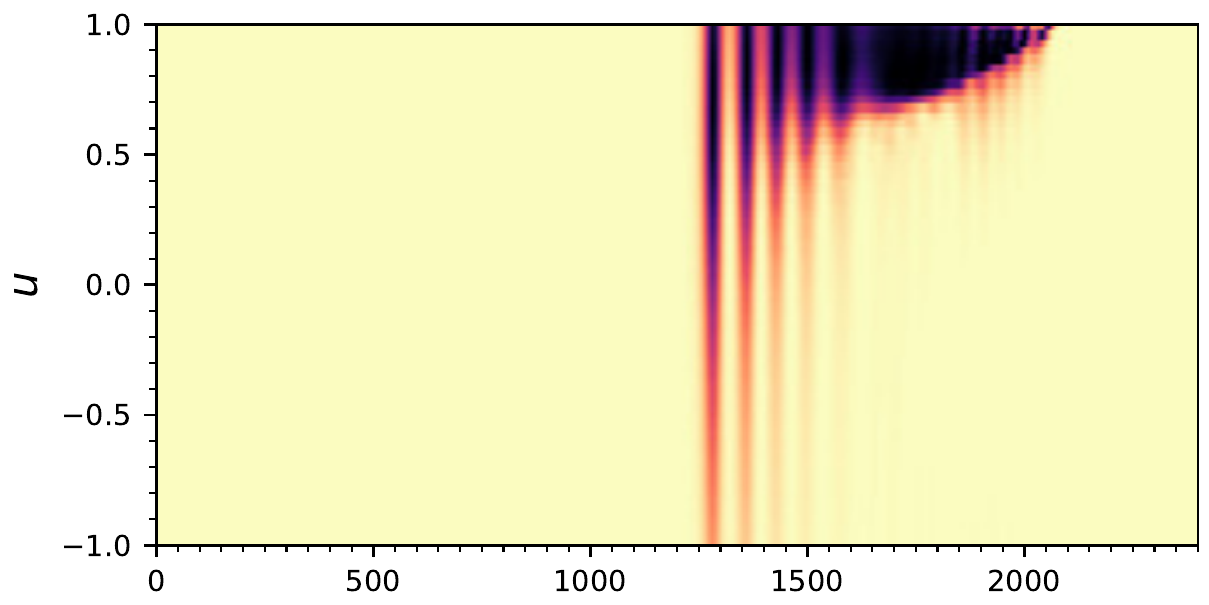} &
      \includegraphics*[scale=\figscale]{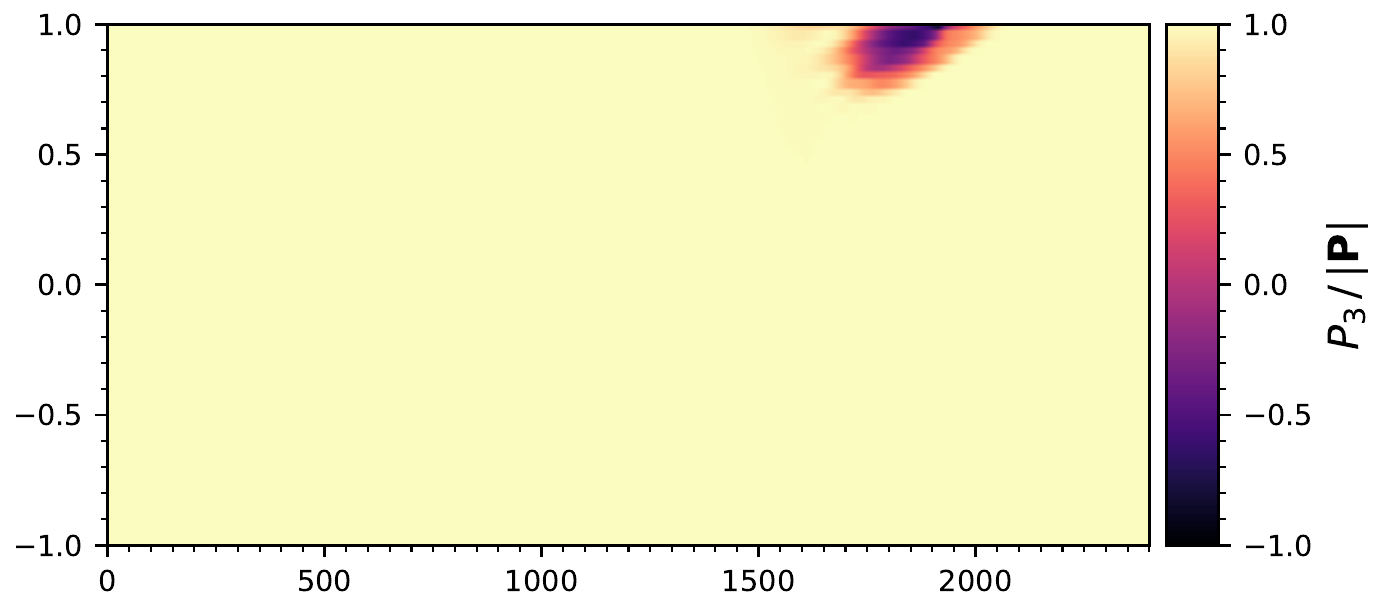} \\
      \includegraphics*[scale=\figscale]{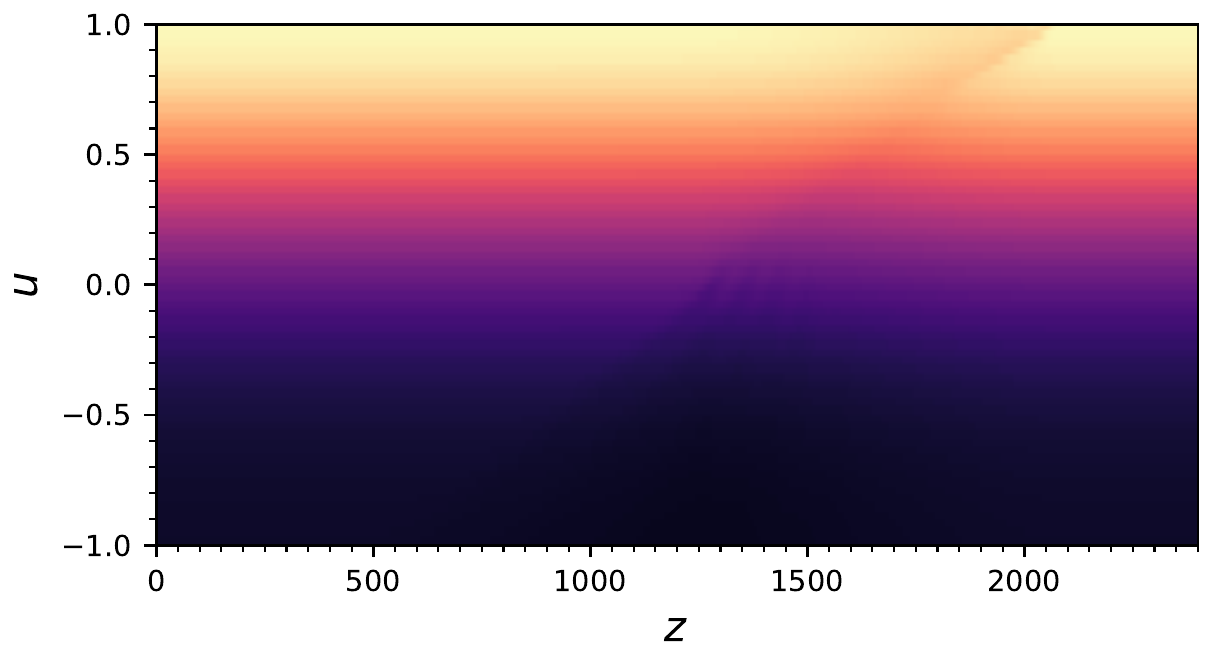} &
      \includegraphics*[scale=\figscale]{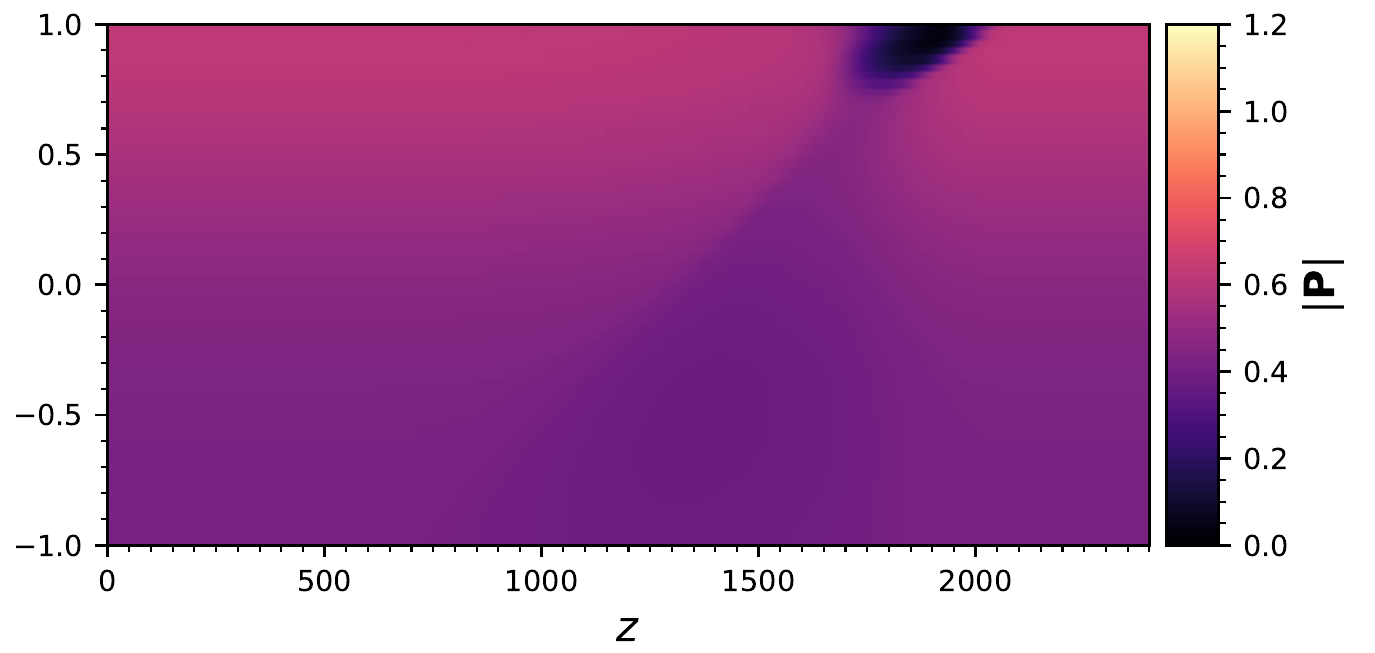} 
    \end{array}$
  \end{center}
  \caption{The polarization vector components $P_1$ (upper panels) and $P_3$ (middle panels), both normalized by the magnitude of the polarization vector $|\bfP|$, and $|\bfP|$ itself (lower panels) of a neutrino gas for all velocity modes at time $t=900$ for the same calculations shown in Fig.~\ref{fig:P-u1}.
  }\label{fig:P-t900}
\end{figure*}

In Fig.~\ref{fig:P-t900} we show the components of the polarization vectors for all velocity modes at $t=900$ in both collision scenarios. The results in the weak collision scenario are again very similar to those in the previous calculation without collisions (see the right panel of Fig.~5 in Ref.~\cite{Martin:2019gxb}). The oscillation wave involves all the velocity modes, but the flavor conversion concentrates in the forward velocity modes where the ELN crossing occurs. In contrast to the extended flavor conversion region in the weak collision scenario, the flavor conversion in the strong collision scenario exists only in a small region away from where the oscillation wave was originally spawned and is even more concentrated in the forward direction. Because most of the neutrino modes do not experience significant flavor conversion, the overall ELN density remains largely homogeneous in the strong collision scenario.

In this work we have adopted a convention different from that in Ref.~\cite{Martin:2019gxb} by normalizing the integral of initial $P_3$ at $z=0$ instead of individual polarization vectors (see Eq.~\ref{eq:P-norm}). Because of the setup of the calculation, $|\bfP_u(t,z)|$ is approximately described by the solid curve in the upper panel of Fig.~\ref{fig:damping} at $t=0$ and for all $z$. In the lower panels of Fig.~\ref{fig:P-t900} we show the magnitudes of the polarization vectors at $t=900$. In the weak collision scenario, $|\bfP_u|$ remains approximately homogeneous even at $t=900$, although there is some minor changes due to the collisions. In the strong collision scenario, however, $|\bfP_u|$ is not only homogeneous but also nearly isotropic. It has the smallest magnitude for the neutrino field that has the largest flavor conversion, which is yet another reason why the neutrino flavor conversion in the strong collision scenario does not change the ELN density very much. Because the collision term included in our calculations changes the direction but not the flavor of the neutrino, and because most of the neutrinos and antineutrinos remain in the initial weak-interaction states in the strong collision scenario, the collisions do not result in a significant flavor depolarization except for the neutrinos and antineutrinos with notable flavor conversions.

\begin{figure*}[htb!]
  \begin{center}
    $\begin{array}{@{}l@{\hspace{0.01in}}l@{\hspace{0.01in}}l@{}}
      \includegraphics*[scale=\figscale]{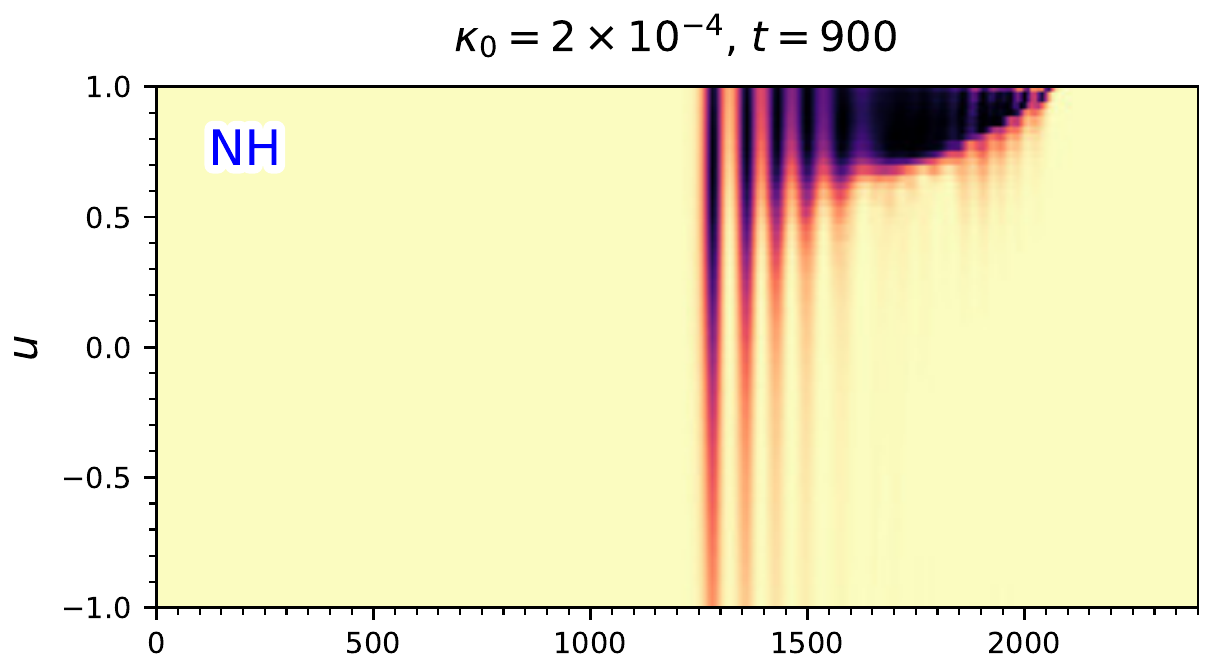} &
      \includegraphics*[scale=\figscale]{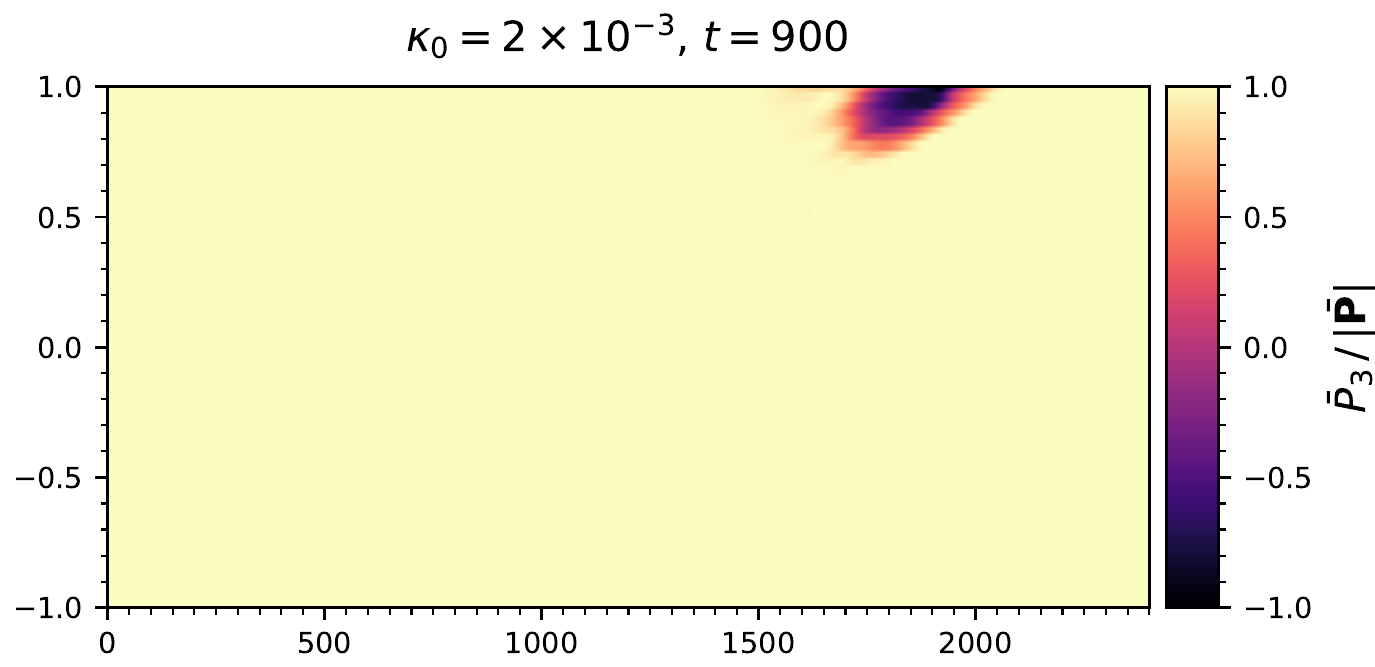} \\
      \includegraphics*[scale=\figscale]{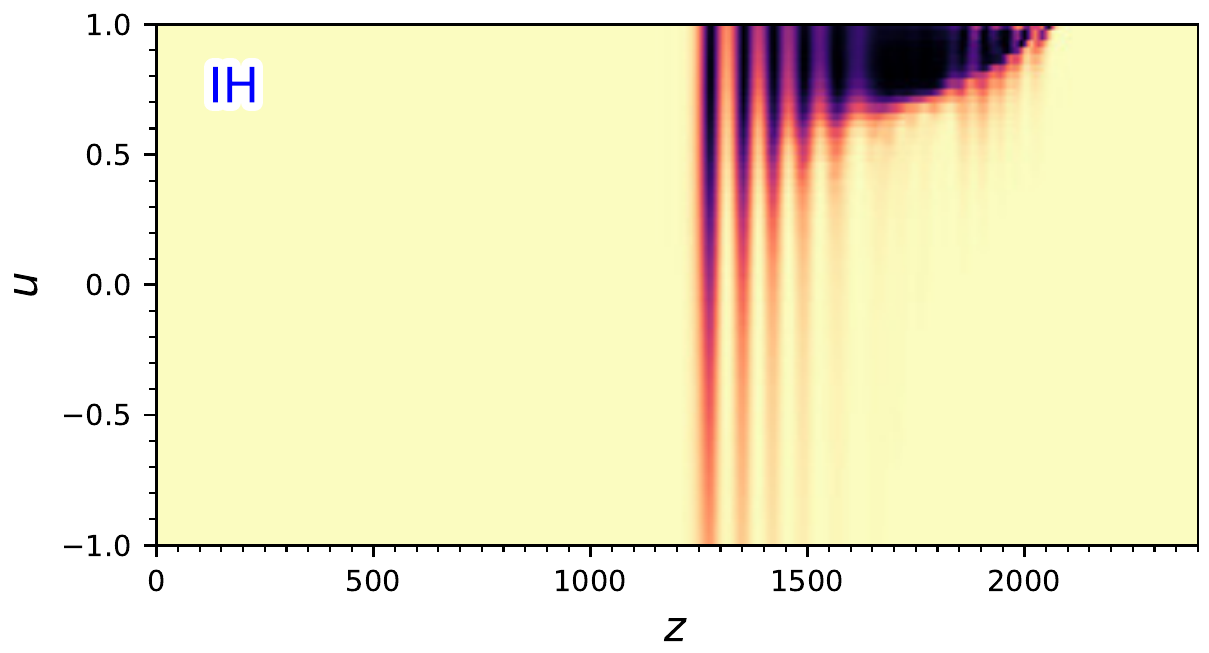} &
      \includegraphics*[scale=\figscale]{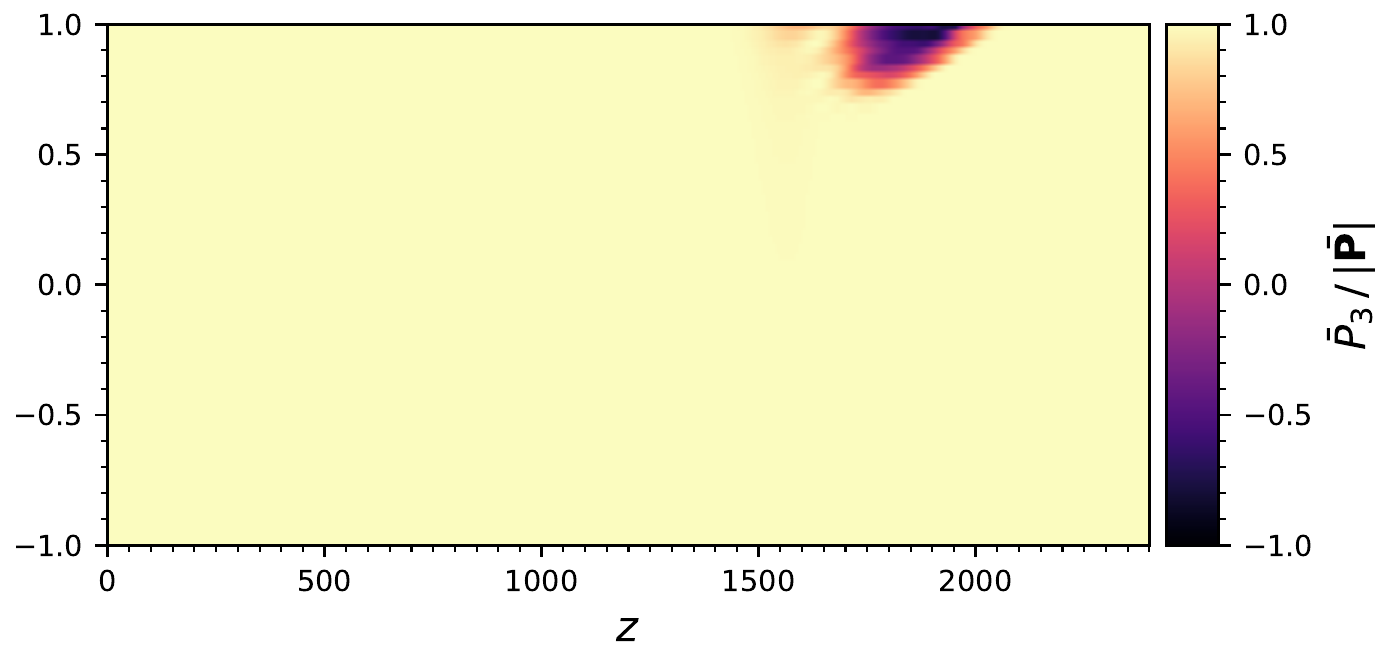}
    \end{array}$
  \end{center}
  \caption{The polarization vector components $\Pa_3$ of the antineutrinos (normalized by the magnitude of the vector $|\bfPa|$) in the calculations shown in Fig.~\ref{fig:P-t900} that employs the normal neutrino mass hierarchy (NH, upper panels) and those in the similar calculations with the inverted neutrino mass hierarchy (IH, lower panels), respectively. 
  }\label{fig:Pa3-t900}
\end{figure*}

In Fig.~\ref{fig:Pa3-t900} we show $\Pa_3$ at $t=900$ in the calculations that employ the NH and IH calculations, respectively. Because an antineutrino oscillates in the same way as a neutrino with the negative energy \cite{Duan:2005cp}, the similarity between these plots and those for $P_3$ in Fig.~\ref{fig:P-t900} shows the insensitiveness of fast oscillation waves to the neutrino energies. Fig.~\ref{fig:Pa3-t900} also demonstrates that the neutrino mass hierarchy has very little impact on fast oscillation waves which confirms the validity of neglecting the vacuum term in most of the literature.

\section{Discussion and conclusions}\label{sec:discussion}

We have performed numerical calculations of fast oscillations in dense neutrino media using mono-energetic neutrinos and antineutrinos with $|\omega|/\mu=10^{-5}$, a typical value on the neutrino sphere inside a core-collapse supernova and for the atmospheric mass-splitting, where $\omega$ and $\mu$ are the effective vacuum oscillation frequency of the neutrinos and the strength of the coherent neutrino-neutrino forward scattering, respectively. We found that the mass splitting and the mass ordering of the neutrino have very little effect on fast oscillations in our calculations. This result seems to contradict that of Ref.~\cite{Shalgar:2020xns} which suggests that the neutrino energy plays a prominent role in fast flavor conversions. We also did not observe the fine angular structures in neutrino flavor conversions as in Ref.~\cite{Shalgar:2020xns}. This discrepancy may be because Ref.~\cite{Shalgar:2020xns} studied homogeneous neutrino gases which can oscillate only in the $K=0$ modes instead of producing a propagating oscillation wave. The stability analysis also shows that the homogeneous modes have small instabilities in our models (see Fig.~\ref{fig:DR}). In any case, it seems unlikely that the neutrino mass splitting can have a big impact on fast oscillation waves where the neutrino vacuum oscillation frequencies are much smaller than the exponential growth rates of their flavor instabilities.

We have also investigated the effects of the elastic neutrino-nucleon collisions on fast oscillation waves. Although collisions may trigger fast oscillations \cite{Capozzi:2018clo}, we found that they damp the flavor oscillation waves on the scale of the neutrino mean free path $\kappa_0^{-1}$. We have shown analytically that the collisions make the neutrino angular distributions significantly more isotropic and thus diminish or even kill the ELN crossing completely after the neutrinos have traveled for one mean free path. In addition, the collisions also damp of the neutrino oscillation waves by adding imaginary components to the collective oscillation frequencies. Because collisions mix the flavor contents of different neutrino momentum modes, they tend to reduce the magnitude of the polarization vectors and cause flavor depolarization when the neutrinos of different momenta have flavor evolution histories. Our findings contradict the results of Ref.~\cite{Shalgar:2020wcx} which suggest that collisions can enhance fast flavor conversions in homogeneous neutrino gases on time scales much larger than $\kappa_0^{-1}$.

In the strong collision scenario, we have found that the neutrino oscillation wave continues to propagate even after the ELN crossing has faded away. This finding should not come as a surprise because the dispersion relation still exists for the neutrino oscillation wave even without the ELN crossing. It is, therefore, entirely possible that fast neutrino oscillation waves that are produced inside the proto-neutron star where ELN crossings exist (see, e.g., Ref.~\cite{Glas:2019ijo}) can propagate through the regions without the ELN crossing and influence various aspects of the supernova physics.

Although we have assumed that the neutrinos are mono-energetic and that the collisions only change the directions of the neutrinos, we do not expect our results will be significantly modified when these constraints are relaxed. This is because, as explained above, the neutrino mass splitting does not have a big impact on fast oscillation waves, and neutrinos and antineutrinos of different energies will have the same flavor evolution in the limit of a vanishing neutrino mass splitting. 

We have used unrealistically large neutrino collision rates in this work to illustrate their impacts. The typical mean free path on the surface of the proto-neutron star is about $\kappa_0^{-1}\sim 10\text{ km}$, which gives $\kappa_0/\mu\sim 10^{-6}$. This is orders of magnitude smaller than the value we used even in the weak collision scenario. Our calculations suggest that fast neutrino oscillation waves, once produced near or inside the surface of the proto-neutron star, can propagate almost unimpeded outward as the neutrinos become essentially free streaming outside the neutrino sphere. 

In this work we have limited the time ranges of the calculations to avoid the unphysical consequences of the periodic boundary condition. It is possible that small angular structures may develop on longer time scales as they do in homogeneous models (see, e.g., Ref.~\cite{Johns:2020qsk}). We have also assumed an arbitrary ELN distribution which favors the production of fast oscillation waves. It remains to be seen how the absorptions, emissions, and collisions of neutrinos that produce the ELN crossing in the first place may interplay with the production and propagation of the fast neutrino oscillation waves. 

\section*{Acknowledgments}
We thank S.~Abbar for the useful discussion. The work of J.~D.~M.\ and H.~D.\ is supported by the US DOE
NP grant No.\ DE-SC0017803 at UNM. 
The work of J.~C.\ and V.~C.\ is supported by US DOE through LANL. LANL is operated by Triad National Security, LLC, for  NNSA of US DOE (Contract No. 89233218CNA000001). 
The work of J.~C.\ is also supported by the LANL LDRD program. The work of J.~D.~M.\ is also supported by the US DOE Office of Science Graduate Student (SCGSR) program and DOE NP under Contract No.\ DE-AC52-06NA25396. The SCGSR program is administered by the Oak Ridge Institute for Science and Education (ORISE) for the DOE. ORISE is managed by ORAU under Contract No. DE-SC0014664. All opinions expressed in this paper are the authors’ and do not necessarily reflect the policies and views of DOE, ORAU, or ORISE.
This research used resources of NERSC, a US DOE Office of Science User Facility located at LBNL, operated under Contract No.\ DE-AC02-05CH11231.

\bibliography{tlc}

\end{document}